\documentclass[useAMS,usenatbib]{mn2e}
\usepackage{aas_macros,graphicx,times,multirow}

\usepackage{graphicx}
\usepackage{standalone}
\usepackage{subfig}
\usepackage{multirow}
\usepackage{array}

\newcommand{\xmm}{{\em XMM-Newton}}
\newcommand{\chan}{{\em Chandra}}

\title[GTC observations of $\gamma$-ray pulsars]{Observations of three young $\gamma$-ray pulsars with the Gran Telescopio Canarias}
\author[R. P. Mignani, N. Rea,  V. Testa, et al. ]
{\parbox{\textwidth}{R. P. Mignani$^{1,2}$\thanks{E-mail: mignani@iasf-milano.inaf.it}, 
N. Rea$^{3,4}$, 
V. Testa$^{5}$,
M. Marelli$^{1}$,
A. De Luca$^{1,6}$,
M. Pierbattista$^{7}$,
A. Shearer$^{8}$,
D. F. Torres$^{4}$,
E. De O\~{n}a Wilhelmi$^{4}$
} 
\\ \\
$^{1}$ INAF - Istituto di Astrofisica Spaziale e Fisica Cosmica Milano, via E. Bassini 15, 20133, Milano, Italy\\
$^{2}$ Janusz Gil Institute of Astronomy, University of Zielona G\'ora, Lubuska 2, 65-265, Zielona G\'ora, Poland \\
$^{3}$ Anton Pannekoek Institute for Astronomy, University of Amsterdam, Postbus 94249, NL-1090 GE Amsterdam, the Netherlands \\
$^{4}$ Instituto de Ciencias de lÕEspacio (ICE, CSIC-IEEC), Campus UAB, Carrer Can Magrans s/n, 08193 Barcelona, Spain \\
$^{5}$ INAF - Osservatorio Astronomico di Roma, via Frascati 33, 00040, Monteporzio, Italy \\
$^{6}$ INFN - Istituto Nazionale di Fisica Nucleare, sezione di Pavia, via A. Bassi 6, 27100, Pavia, Italy \\
$^{7}$ Maria Curie-Sklodowska University, Department of Astrophysics and Theory of Gravity, ulica Radziszewskiego 10, 20-031 Lublin, Poland\\
$^{8}$ Centre for Astronomy, National University of Ireland, Newcastle Road, Galway, Ireland \\
}

\begin{document}

\date{Accepted 1988 December 15. Received 1988 December 14; in original form 1988 October 11}

\pagerange{\pageref{firstpage}--\pageref{lastpage}} \pubyear{2002}

\maketitle

\label{firstpage}

\begin{abstract}
We report the analysis of the first deep optical observations of three isolated $\gamma$-ray pulsars detected by the  {\em Fermi Gamma-ray Space Telescope}: the radio-loud PSR\, J0248+6021 and PSR\, J0631+1036, and the radio-quiet PSR\, J0633+0632.  The latter  has also been detected in the X rays.  The pulsars are very similar in their 
 spin-down age ($\tau \sim$40--60 kyrs),
 spin-down energy ($\dot{E} \sim10^{35}$ erg s$^{-1}$), and dipolar surface magnetic field ($B \sim 3$--$5\times10^{12}$ G).
These pulsars are promising targets for multi-wavelength observations, since they have been already 
detected in $\gamma$ rays and in radio or X-rays. None of them has been detected yet in the optical band.
We observed the three pulsar fields in 2014 with the Spanish 10.4m Gran Telescopio Canarias (GTC).  We could not find any candidate optical counterpart to the three pulsars close to their most recent radio or {\em Chandra} positions down to $3 \sigma$ limits of $g'\sim27.3$,  $g'\sim27$,  $g'\sim27.3$ for PSR\, J0248+6021, J0631+1036, and J0633+0632, respectively. From the inferred optical upper limits and estimated distance and interstellar extinction, we derived limits on the pulsar optical luminosity. We also searched for the X-ray counterpart to PSR\, J0248+6021 with \chan\ but we did not detect the pulsar down to a 3$\sigma$ flux limit of $5 \times 10^{-14}$ erg cm$^{-2}$ s$^{-1}$ (0.3--10 keV).  For all these pulsars, we compared the optical flux upper limits with the extrapolations in the optical domain of the $\gamma$-ray spectra
and compared their multi-wavelength properties with those of other $\gamma$-ray pulsars of comparable age. 
\end{abstract}


\section{Introduction}

The launch of the {\em Fermi Gamma-ray Space Telescope} in 2008 marked  a revolution in  pulsar $\gamma$-ray studies (see, e.g. Caraveo 2014; Grenier \& Hardings 2015 for recent reviews), thanks to the unprecedented performances of its {\em Large Area Telescope} ({\em LAT};  Atwood et al.\ 2009). At the time of writing, the public catalogue of {\em Fermi} $\gamma$-ray pulsars\footnote{{\texttt https://confluence.slac.stanford.edu/display/GLAMCOG/\\Public+List+of+LAT-Detected+Gamma-Ray+Pulsars}} includes 205 objects, compared to the seven detected by the {\em Compton Gamma-ray Observatory (CGRO)} (e.g. Thompson 2008).   Most of them are isolated, i.e. not in binary systems. Interestingly, about 30\% of these $\gamma$-ray pulsars have no radio counterpart, despite deep radio searches, and are dubbed radio-quiet. Therefore, this  unprecedentedly large and diverse sample provides us with an unique opportunity to characterise  the pulsar multi-wavelength  spectra and emission properties over several energy decades, from the optical to the $\gamma$-rays, possibly spotting differences between different pulsar populations (e.g. Marelli et al.\ 2015). This is key to understand both the physics of the complex radiation processes in pulsar magnetospheres 
and the behaviour of relativistic particles and radiation under extreme magnetic field conditions. To this aim, building an as wide as possible multi-wavelength observational data base is essential. 

While X-ray observations have been obtained for about half of the isolated {\em Fermi} pulsars in the Second Catalogue of {\em Fermi} Gamma-ray Pulsars (2PC; Abdo et al.\ 2013), albeit at different levels of statistics, the optical coverage is still sparse, mainly owing to the intrinsic neutron star faintness, relatively large distances, high (or uncertain) interstellar extinction, field crowding at low Galactic latitudes, and the lack of sensitive observations with 10m-class telescopes.  As a matter of fact, only very few isolated {\em Fermi} pulsars have been detected
both in the X rays and in the optical. These are the Crab (PSR\, B0531+21) and Vela (PSR\, B0833$-$45) pulsars, PSR\, B1509$-$58, PSR\, B0656+14, PSR\, B1055$-$52 and Geminga (see Abdo et al.\ 2013 and references therein), all detected as $\gamma$-ray pulsars before the launch of {\em Fermi} (Thompson 2008) and identified in the optical within a few years after their discovery.  Among the new $\gamma$-ray pulsars discovered by {\em Fermi}, a candidate optical counterpart has been identified for PSR\, J0205+6449 (Moran et al.\ 2013), whereas for PSR\, B0540$-$69 in the Large Magellanic Cloud, only recently detected as a $\gamma$-ray pulsar  
(Ackermann et al.\ 2015), the optical counterpart 
has been known since the early 1990s (Caraveo et al.\ 1992), following its discovery as an X-ray and optical pulsar (Seward et al.\ 1984; Middleditch \& Pennypacker 1985).
Another handful of $\gamma$-ray pulsars have been observed after their discovery by {\em Fermi}, but not detected yet, with 10m-class telescopes:  PSR\, J1357$-$6429 (Mignani et al.\ 2011), PSR\, J1028$-$5819 (Mignani et al.\ 2012), PSR\, J1048$-$5832 (Razzano et al.\  2013; Danilenko et al.\ 2013), PSR\, J0007+7303 (Mignani et al.\ 2013),  PSR\,  J0357+3205 (De Luca et al.\ 2011;  Kirichenko et al.\ 2014), and PSR\, J2021+3651 (Kirichenko et al.\ 2015). Recently, PSR\, J1357$-$6429 might have been identified in the near infrared (Zyuzin et al.\ 2016).

  \begin{table*}
\begin{center}
\caption{Coordinates, reference epoch, and spin-down parameters of the {\em Fermi} pulsars discussed in this work, collected from the ATNF pulsar data base (Manchester et al.\ 2005).}
\label{psr}
\begin{tabular}{lllcccccc} \hline
Pulsar 				     &$\alpha_{J2000} $	& $\delta_{J2000}$&    Epoch & P$_{\rm s}$	&    $\dot{P_{\rm s}}$ 	& $\tau$ & B & $\dot{E}$ \\ 
                                          &    $^{(hms)}$   &      $^{(\circ ~'~")}$   &  MJD &   (s) &  (10$^{-14}$s s$^{-1}$)    &  ($10^{4}$ yr) & ($10^{12}$ G)  & ($10^{35}$ erg cm$^{-2}$ s$^{-1}$)  \\ \hline
J0248+6021$^1$           &  02 48 18.617(1) &  +60 21 34.72(1) & 54000&  0.217 & 5.51 & 6.24 & 3.5 &  2.1 \\
J0631+1036           &  06 31 27.524(4) &  +10 37 02.5(3)      & 53850 & 0.287 & 0.10 & 4.36 & 5.55 & 1.7 \\
J0633+0632           &  06 33 44.21(2)   &  +06 32 34.9(1.6)    & 54945& 0.297 & 7.95 & 5.92 & 4.92 & 1.2 \\ \hline  
\end{tabular}
\end{center}
$^1$ The pulsar has a proper motion $\mu_{\alpha} cos(\delta) =48 \pm 10$ mas yr$^{-1}$; $\mu_{\delta}=48\pm4$ mas yr$^{-1}$ (Theureau et al.\ 2011).  The spin period derivative $\dot{P_{\rm s}}$ and values inferred from it have been corrected for the Shklovskii  effect.
\end{table*}

Here, we present the first deep optical observations of a group of {\em Fermi} pulsars: PSR\, J0248+6021, PSR\, J0631+1036, and PSR\, J0633+0632 (Abdo et al.\ 2010a; 2013), carried out with the Spanish 10.4m Gran Telescopio Canarias (GTC) at the  La Palma Observatory (Roque de Los Muchachos, Canary Islands, Spain), as a part of a larger program aimed at surveying {\em Fermi} pulsars in the northern hemisphere.
 A parallel program 
  in the southern hemisphere is being carried out with the ESO's Very Large Telescope and the results will be discussed in a companion paper (Mignani et al., in preparation). 
These three pulsars have 
spin-down ages $\tau \sim$ 43.6--62.4 kyrs, 
spin-down energies $\dot{E} \sim$(1--2)$\times 10^{35}$ erg s$^{-1}$ and dipolar surface magnetic fields $B \sim 3$--$5\times10^{12}$ G (see Table 1).

PSR\, J0248+6021 (spin period P$_{\rm s}$=0.217 s) was discovered in radio during a northern Galactic plane survey with the Nancay radio telescope (Theureau et al.\ 2011). The pulsar X-ray counterpart was not detected by  {\em Swift} and {\em Suzaku} (Marelli et al.\ 2011) and the field has not been observed by \xmm. Only one short (10 ks) observation has been obtained with \chan\  (Obs ID 13289; PI G. Garmire)  but the pulsar was not detected (Prinz \& Becker 2015).
A quick follow-up observation using optical data from the Ultra-Violet/Optical Telescope (UVOT; Roming et al.\ 2005) aboard {\em Swift} did not show any object at the best-fit radio timing position of the pulsar.  The PSR\, J0248+6021 field was observed in H$_{\alpha}$ (Brownsberger \& Romani 2014) to search for a bow-shock produced by the pulsar motion in the interstellar medium but neither extended nor point-like emission associated with the pulsar was detected. 

PSR\, J0631+1036 (P$_{\rm s}$ = 0.287 s) was detected during a radio follow-up of unidentified {\em Einstein} X ray-sources  (Zepka et al.\ 1996).  The pulsar is yet undetected in X rays (Kennea et al.\ 2002; Marelli et al.\ 2011). A tentative identification of an X-ray counterpart by both {\em ROSAT}  (Zepke et al.\ 1996) and {\em ASCA} (Torii et al.\ 2001) was found to be the result of a  spurious association.
In $\gamma$-rays, a marginal evidence of pulsations was found in the {\em CGRO} data  (Zepka et al.\ 1996) and was, later,  confirmed by {\em Fermi} (Weltevrede et al.\  2010). A search for optical pulsations from the undetected pulsar counterpart (Carrami\~{n}ana et al.\ 2005) was carried out with negative results. The field was also observed in a targeted observation with the 2.5m Isaac Newton Telescope (INT) at the La Palma Observatory (Collins et al.\ 2011) as part of a pilot survey of {\em Fermi} pulsar fields under the International Time Proposal ITP02 (PI. A. Shearer), 
but the pulsar was undetected. 
PSR\, J0631+1036 was also not detected in the pulsar H$_{\alpha}$  survey of Brownsberger \& Romani (2014). 

PSR\, J0633+0632 (P$_{\rm s}$ = 0.297 s) is a radio-quiet pulsar and one of the first $\gamma$-ray pulsars detected through a blind search in the {\em Fermi} data (Abdo et al.\ 2009a). Deep searches at 34 MHz (Maan \& Aswathappa 2014) confirmed that the source is undetected also at long radio wavelength. After a preliminary detection by {\em Swift} (Abdo et al.\ 2009a), PSR\, J0633+0632 has been detected in X rays by both {\em Suzaku} (Marelli et al.\ 2011) and  {\em Chandra} (Ray et al.\ 2011), which also found evidence of a faint arcminute-long  pulsar wind nebula (PWN) south of the pulsar. No X-ray pulsations from PSR\, J0633+0632 have been detected yet. In the optical, the field was observed for the first time with the INT (Collins et al.\ 2011) but no counterpart to the pulsar was detected.

This manuscript is organised as follows: observations, data reduction and analyses are described in Sectn. 2, while the results are presented and discussed in Sectn. 3. and 4, respectively. Conclusions follow.

\section{Observations and data reduction}

\subsection{GTC Observations}

We obtained deep observations of the pulsar fields with the GTC between December 18, 2014 and January 14, 2015 under   programme GTC23-14B (PI. N. Rea).  The observations were performed in service mode  with the Optical System for Imaging and low Resolution Integrated Spectroscopy (OSIRIS). The instrument is equipped with a two-chip E2V CCD detector with a nominal field--of--view (FoV) of $7\farcm8\times8\farcm5$ that is actually decreased to $7\farcm8 \times 7\farcm8$ due to the vignetting of Chip 1. The pixel size of the CCD is 0\farcs25 ($2\times2$ binning). In total, we took a minimum of three sequences of 5 exposures in the Sloan $g'$  ($\lambda=4815$ \AA; $\Delta \lambda=1530$\AA) $r'$  band ($\lambda=6410$ \AA; $\Delta \lambda=1760$\AA) filters with exposure time of 155 s,  to minimise the saturation of bright stars in the field and remove cosmic ray hits. Each sequence was repeated twice per each filter and per each target. Exposures were dithered by 20\arcsec\ steps in right ascension and declination.
In all cases, the targets were positioned at the nominal aim point in Chip 2. Observations were performed in dark time and clear sky conditions, with seeing mostly below 1\farcs0 and the targets close to the zenith. The journal of the GTC observations is summarised in Table \ref{gtc}. Short (0.5--3 s) exposures of the field of the standard star PG\, 2336+004B (Smith et al.\ 2002) were also acquired each night for photometric calibration and zero point trending\footnote{{\texttt www.gtc.iac.es/instruments/osiris/media/zeropoints.html}}, together with twilight sky flat fields, as part of the OSIRIS service mode calibration plan (Cabrera-Lavers et al.\ 2014).
 We reduced our data  (bias  subtraction, flat-field correction) using standard tools in the {\sc IRAF}\footnote{IRAF is distributed by the National Optical Astronomy Observatories, which are operated by the Association of Universities for Research in Astronomy, Inc., under cooperative agreement with the National Science Foundation.} package {\sc ccdred}. Single dithered exposures were then aligned, average-stacked, and filtered by cosmic rays using the task {\tt drizzle}.
We 
adopted the standard extinction coefficients measured for the La Palma Observatory\footnote{{\texttt  www.ing.iac.es/Astronomy/observing/manuals/ps/tech\_notes/tn031.pdf}}
to apply the airmass correction.

 \subsection{Optical Astrometry}

We computed the astrometry calibration on the GTC images using the {\em wcstools}\footnote{{\texttt http://tdc-www.harvard.edu/wcstools}} suite of programs, matching the sky coordinates of stars selected from the Two Micron All Sky Survey (2MASS) All-Sky Catalog of Point Sources (Skrutskie et al.\ 2006)
  with their pixel coordinates computed by {\em Sextractor} (Bertin \&Arnouts 1996).  After iterating the matching process applying a $\sigma$-clipping selection to filter out obvious mismatches, high-proper motion stars, and false detections, we obtained mean residuals of $\sim 0\farcs2$ in the radial direction, using at least 30 bright, but non-saturated, 
  2MASS stars. Owing to the pixel scale of the OSIRIS images (0\farcs25), the uncertainty on the centroids of the reference stars is negligible.
  To this value we added in quadrature the uncertainty $\sigma_{tr}\la$  0\farcs07 of the image registration  on the 
 2MASS reference frame. This is given by $\sigma_{tr}$=$\sqrt{n/N_{S}}\sigma_{\rm S}$ (e.g., Lattanzi et al.\ 1997), where $N_{S}$ is the number of stars used to compute the astrometric solution, $n$=5 is the number of free parameters in the sky--to--image transformation model (rotation angle, x-offset, y-offset, x-scale, y-scale), $\sigma_{\rm S} \sim 0\farcs2$ is the mean absolute position error of  2MASS (Skrutskie et al.\ 2006) for stars in the magnitude range  $15.5 \le K \le 13$.
 After accounting for the 0\farcs015  uncertainty on the link of 
2MASS to the International Celestial Reference Frame  (Skrutskie et al.\ 2006),
we ended up with an overall accuracy of $\sim$0\farcs2 on our absolute astrometry.

\begin{figure*}
\centering
\begin{tabular}{cc}
\subfloat[J0248+6021]{\includegraphics[width=8cm]{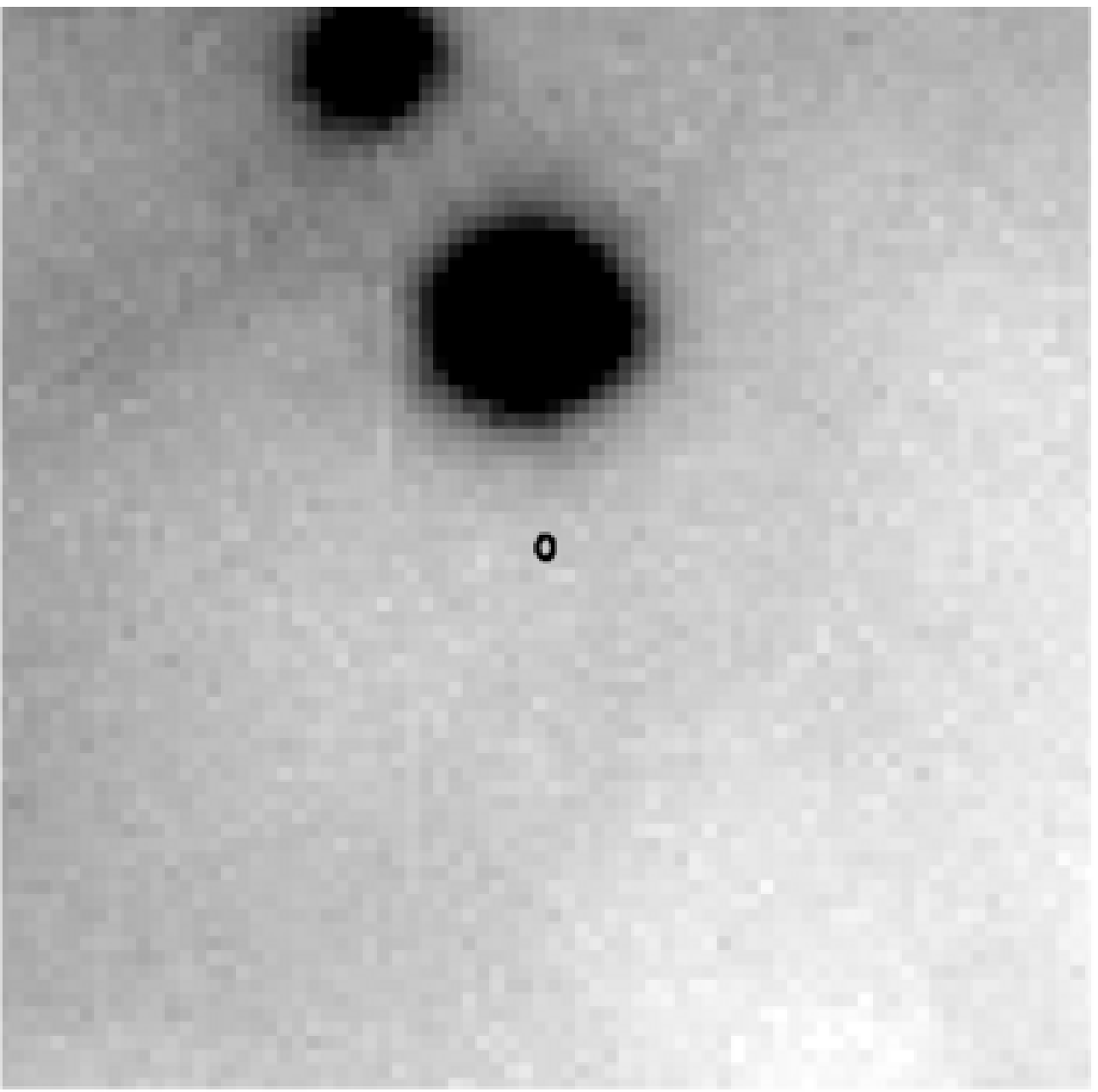}} &
\subfloat[J0631+1036]{\includegraphics[width=8cm]{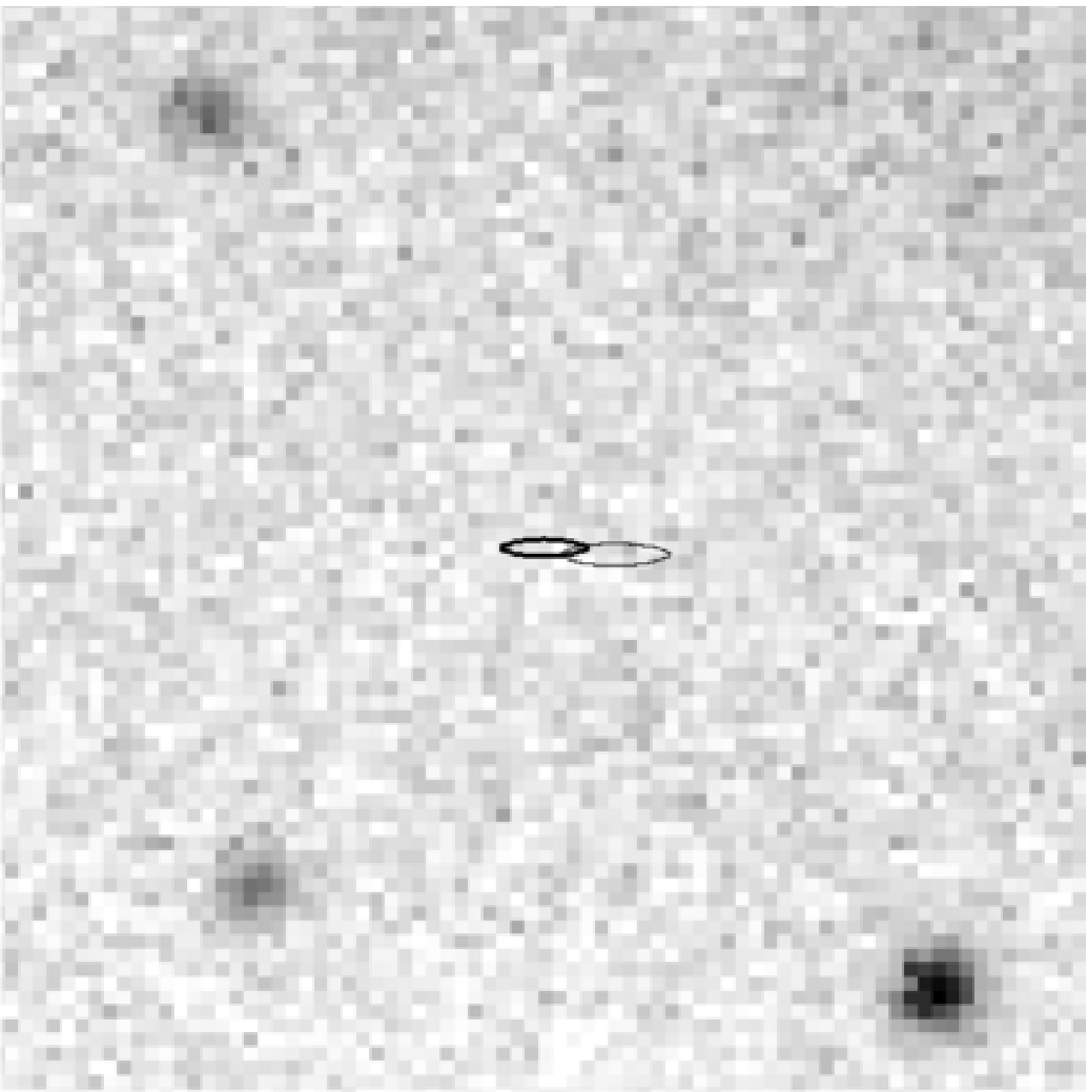}} \\
\subfloat[J0633+0632]{\includegraphics[width=8cm]{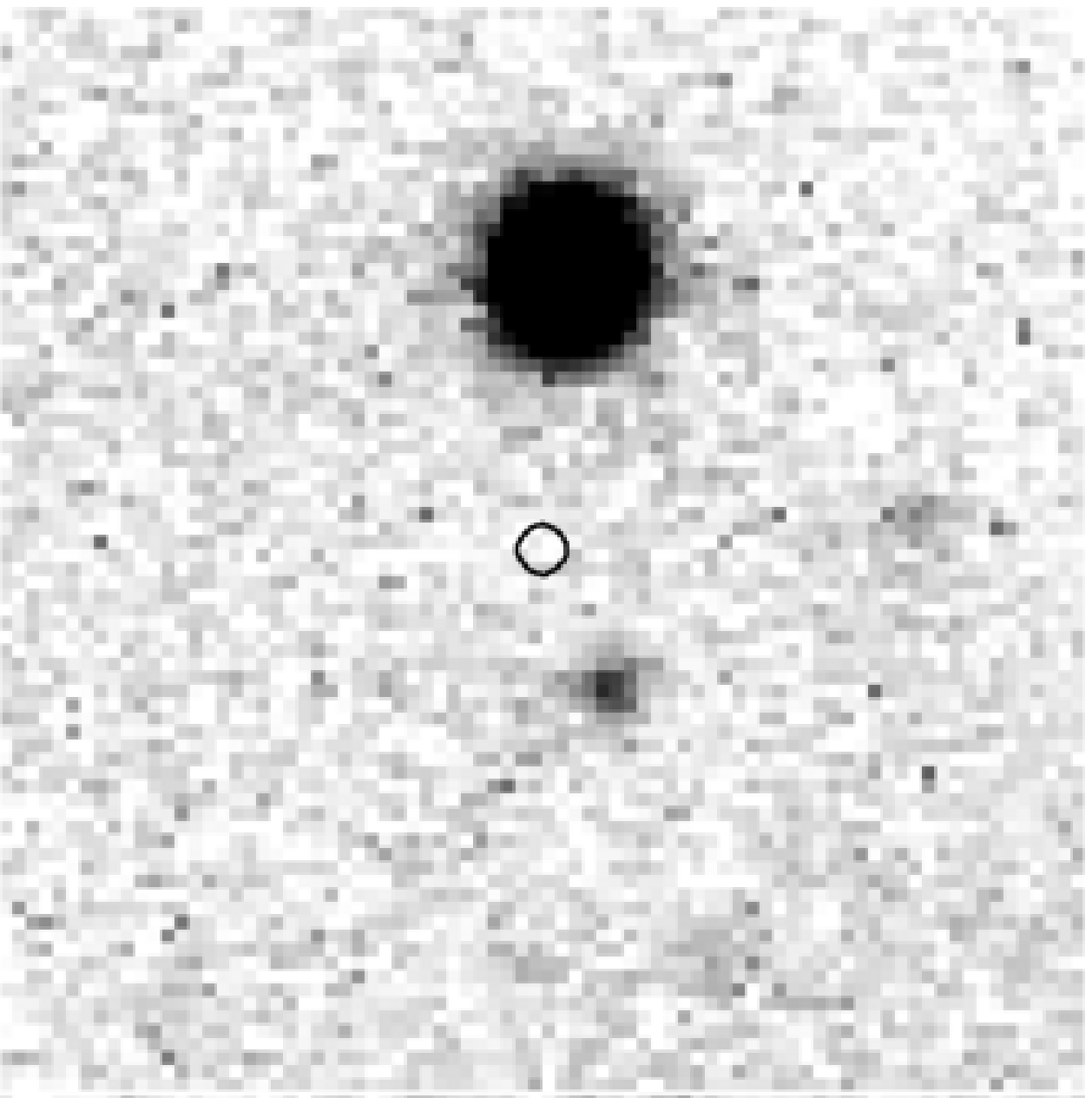}} \\
\end{tabular}
\caption{\label{fc} 
$10\arcsec \times 10\arcsec$ GTC/OSIRIS $r'$-band images of the pulsar fields. North to the top, east to the left. The uncertainties associated with the available radio positions (Sectn.\ 2.4) of PSR\, J0248+6021 and PSR\, J0631+1036 are shown by the ellipses. For the latter, the thick ellipse corresponds to the radio timing position (Yuan et al.\ 2010), whereas the thin ellipse marks the radio timing position of Hobbs et al.\ (2004). For PSR\, J0633+0632, 
the circle indicates the \chan\ position uncertainty  (Ray et al.\ 2011). The size of the ellipses/circle accounts for statistical uncertainties only and not for the systematic uncertainty associated with the astrometry calibration of the OSIRIS images ($\sim$0\farcs2; Sectn.\ 2.2).
}
\end{figure*}

\begin{table}
\centering
\caption{Summary of the GTC optical observations of the three {\em Fermi} pulsars in Table \ref{psr}. Columns list the observing date, band, the total  integration time (T), the average airmass and seeing.}
\label{gtc}
\begin{tabular}{lccccc} \hline
Pulsar & Date      &  Band & T   & airmass & seeing\\
                  & yyyy-mm-dd [MJD]  &    &  (s)  & & (\arcsec)\\ \hline
J0248+6021           & 2015-01-14 [57036]  &  g & 2325 & 1.17  &  0\farcs8\\ 
			                   &   &  g & 2325 & 1.19 & 0\farcs9  \\ 
					     & 2014-12-19 [57010]  &  r & 2625 & 1.29  & 1\farcs1  \\ 
					     &   &  r & 2790 &  1.43 & 1\farcs1  \\ \hline
J0631+1036           & 2014-12-22 [57013]  &  g & 3100 & 1.06 & 1\farcs2 \\ 
				  &  &  r & 2945 & 1.07 &  1\farcs2\\ 
				     & 2014-12-24 [57015]  &  g & 2325 & 1.29 & 1\farcs1 \\ 
					  &  &  r & 2325 & 1.52 &  1\farcs1 \\ \hline
 J0633+0632           & 2014-12-18 [57009] &  g & 2480 & 1.08  & 0\farcs9 \\ 
 				   &   &  g & 2325 & 1.22 & 0\farcs9  \\ 
				  &  &  r & 2325 & 1.13 & 0\farcs8 \\ 
					  &   &  r & 2325 & 1.39  & 0\farcs8 \\ \hline  
\end{tabular}
\vspace{0.5cm}
\end{table}

\subsection{Pulsar coordinate verification}

In order to search for the pulsar optical counterparts, a careful assessment of their coordinates is in order.
To search for the optical counterparts of our pulsars, a first reference is provided by the coordinates listed in the ATNF pulsar data base (Table \ref{psr}). 

For both PSR\, J0248+6021 (Theureau et al.\ 2011) and J0631+1036 these coordinates have been obtained from radio observations. For PSR\, J0248+6021 a radio proper motion has been measured (Theureau et al.\ 2011) and we updated its coordinates  to the epoch of our optical observations (Table \ref{gtc}). 
For PSR\, J0631+1036, the {\em Simbad}\footnote{{\texttt http://simbad.u-strasbg.fr/simbad/}} data base reports coordinates, $\alpha =06^{\rm h}  31^{\rm m} 27\fs540$; $\delta  = +10^\circ 37\arcmin 02\farcs20$, which are only slightly different from those in the ATNF data base (MJD 53850).  Different coordinates  are reported by Hobbs et al.\ (2004):  $\alpha =06^{\rm h}  31^{\rm m} 27\fs516(12)$; $\delta  = +10^\circ 37\arcmin 03\farcs8(9)$ (MJD 51711), obtained by radio timing  observations and possibly affected by glitches and timing noise. However, they are compatible within the errors with those listed in the ATNF data base, obtained from radio timing observations (Yuan et al.\ 2010).
Unfortunately, since PSR\, J0631+1036 is yet undetected in  the X rays (Kennea et al.\ 2002) there is no {\em Chandra} position to compare with. 
For the radio-quiet PSR\, J0633+0632 the coordinates in Table \ref{psr} correspond to the best-fit $\gamma$-ray timing position (Ray et al.\ 2011), which has a relatively large error ($\pm$1\farcs6) in declination. A somewhat different position, however, is obtained from  {\em Chandra} observations (Ray et al.\ 2011):  $\alpha =06^{\rm h}  33^{\rm m} 44\fs143$; $\delta  = +06^\circ 32\arcmin 30\farcs40$ (MJD 55176), with an attached nominal absolute uncertainty of 0\farcs6 at 90\% confidence level\footnote{{\texttt http://cxc.harvard.edu/cal/ASPECT/celmon/}}.  These coordinates are in very good agreement with those obtained from an independent analysis of the same \chan\ data set by Marelli (2012). Strangely enough, the coordinates reported by {\em Simbad}, $\alpha =06^{\rm h}  31^{\rm m} 39\fs4$; $\delta  = +06^\circ 41\arcmin 42\farcs0$, are off by several arc minutes with respect to those reported by Ray et al.\ (2011) and Marelli (2012), and should be ignored.
Hereafter, for PSR\, J0633+0632 we assume the \chan\ coordinates  as a reference.
For PSR\, J0631+1036 and  PSR\, J0633+0632 the maximum time span between the epoch of the reference positions and that our optical observations is 3165 and 1833 days, respectively. Therefore, we allowed for an uncertainty on the reference pulsar positions more generous than the formal one to account for their unknown proper motions. Apart from  PSR\, J0248+6021, the uncertainty on the pulsar position is always larger than the accuracy of our astrometry calibration ($\sim$0\farcs2; Sectn. 2.3).

\section{Results}

\subsection{Source detection and photometry}

Fig. \ref{fc} shows a $10\arcsec\times10\arcsec$ zoom of the GTC $r'$-band images around the pulsar positions. In no case we could find candidate counterparts to the pulsars, which remain, thus, undetected in the optical. For PSR\, J0633+0632, the object closest to the \chan\ position is detected $\sim 2\farcs5$ southwest of it and is obviously unrelated to the pulsar. We computed the $3\sigma$  limiting magnitudes from the standard deviation of the background at the pulsar position (Newberry 1991) estimated in an aperture of diameter equal to the seeing disk (Table \ref{gtc}), after applying the aperture correction. We corrected these values for the atmospheric extinction as described in Sectn. 2.1.  We derived $g'>27.3$,  $r'>25.3$  (PSR\, J0248+6021), $g'>27$,  $r'>26.3$ (PSR\, J0631+1036), $g'>27.3$,  $r'>26.5$ (PSR\, J0633+0632). 
The difference in limiting magnitudes between the $g'$ and $r'$ bands for similar integration times (Table \ref{gtc}) is probably due to the much larger sky brightness in the $r'$-band measured at the Roque de Los Muchachos Observatory\footnote{{\texttt http://www.gtc.iac.es/instruments/osiris/}}.
 For PSR\, J0248+6021, the much shallower limit in the $r'$ band with respect to the $g'$-band one is due both to the worse seeing conditions ($1\farcs1$), which broadens the halo of a relatively bright star $\sim 4\farcs5$ north of the pulsar (Fig. \ref{gtc}, top left), and the higher surface brightness of the W5 HII region, which covers a large part of the OSIRIS field of view. Both effects increase the rms of the background at the pulsar position.
For PSR\, J0631+1036, the detection of the optical counterpart was hampered by the presence of a very bright star (B$\sim$10.4) about 1\farcm5 south of the pulsar position. Unfortunately, although the telescope roll angle and pointing were chosen both to centre the star in the gap between the two CCDs and place the pulsar at a safe distance from the chip edge, the bright wings of the star's PSF still affects the background at the pulsar position, also owing to the non-optimal seeing conditions (up to 1\farcs2; Table \ref{gtc}).

\begin{figure}
\centering
{\includegraphics[height=8.5cm,angle=0,clip=]{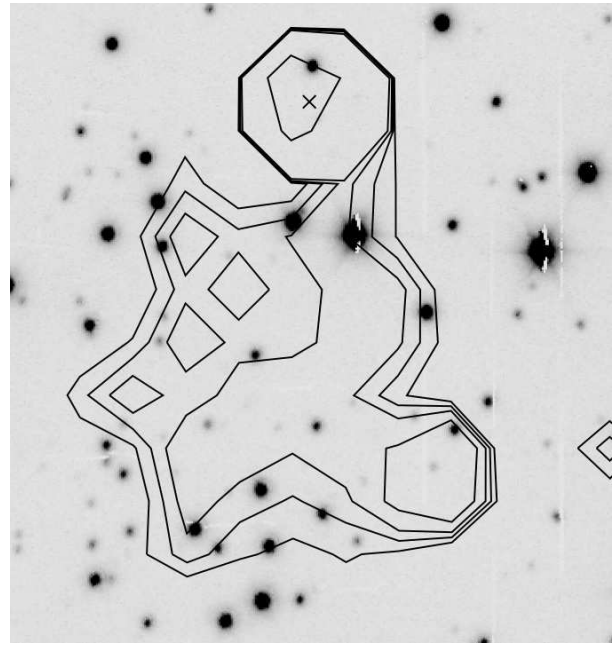}}
\caption{\label{pwn} 
Zoom ($1\farcm5\times1\farcm4$) of the GTC/OSIRIS image of the PSR\, J0633+0632  field  ($r'$ band). North to the top, east to the left. The X-ray contour map from the \chan\ observation (Ray et al.\ 2011) is overlaid in black. For a better representation, the \chan\ image has been smoothed with a Gaussian function using a Kernel radius of 3 pixels. The X-ray contours at the top correspond to the emission maximum at the pulsar position (marked by the cross), whereas the ones at the bottom correspond to the PWN. The emission maximum southwest of the pulsar is associated with a point-like X-ray source unrelated to the PWN (Ray et al.\ 2011).}
\end{figure}

We also used the GTC images to search for a possible evidence of extended optical emission around  PSR\, J0633+0632, which could be associated with the PWN observed in X rays by {\em Chandra} (Ray et al.\ 2011).  A section of the GTC $r'$-band image encompassing the full PWN field is shown in Fig.\ref{pwn}. As seen, no evidence of diffuse emission is found in the region corresponding to the spatial extent of the PWN.  This confirms that PWNe are challenging targets in the optical. Indeed, optical and/or infrared PWNe have been detected so far only around five $\gamma$-ray pulsars: the Crab (Hester 2008), PSR\, B0540$-$69 (e.g., Mignani et al.\ 2010), PSR\, J0205+6449 (Shibanov et al.\ 2008; Slane et al.\ 2008), PSR\, J1124$-$5916 (Zharikov et al.\ 2008), and PSR\, J1833$-$1034 (Zajczyk et al.\ 2012), with the pulsar counterpart being unresolved from the PWN in the last two cases. In the X rays, about 100 PWNe or candidates have been detected so far (e.g., Kargaltsev et al.\ 2015).  We set a $3\sigma$ upper limit of  $\sim 27.7$ and $\sim 26.8$ magnitudes arcsec$^{-2}$ on the surface brightness of the PWN in the $g'$ and $r'$-bands, respectively. These limits have been computed from the standard deviation of the background estimated from star-free regions across the PWN area.  

\subsection{X-ray data analysis}

Two of the three $\gamma$-ray pulsars in Table \ref{psr} are still undetected in the X rays. No new X-ray observations exist for PSR\, J0631+1036, whereas for PSR\, J0248+6021 we found \chan\ data  (Obs ID 13289; PI G. Garmire), which have been recently analysed by Prinz \& Becker (2015). The observation was performed on September 25 2012 (01:57:45 UT) with the Advanced CCD Imaging Spectrometer (ACIS) in  the {\em FAINT} data mode for an effective exposure time of 9.22 ks. The target was placed at the nominal aim position at centre of the ACIS-I detector.  We downloaded the data from the Science Archive\footnote{{\texttt http://cxc.harvard.edu/cda/}} and analysed them with the \chan\ Interactive Analysis of Observations Software (CIAO) v4.6\footnote{{\texttt http://cxc.harvard.edu/ciao/index.html}}. Firstly, we reprocessed level 1 files using the standard {\tt chandra\_repro} script. Then, we retrieved counts image, flux-calibrated image and exposure map in the 0.3--10 keV energy band using the standard {\tt fluximage} script. Taking into account the point spread function spatial distribution, we ran the source detection using the {\tt wavdetect} task. We did not found any source (at a 3$\sigma$ level) positionally consistent with the pulsar position, corrected for its proper motion at the epoch of the \chan\ observation (MJD=56195). Thus, we confirm the non-detection of PSR\, J0248+6021, as reported by Prinz \& Becker (2015). We evaluated the flux upper limit assuming an absorbed power-law
spectrum with a photon index $\Gamma_{\rm X}=2$ and an hydrogen column density $N_{\rm H}= 0.8 \times 10^{22}$ cm$^{-2}$, as assumed in Abdo et al.\ (2013) and set to the Galactic value for the pulsar direction obtained with {\tt Webtools}\footnote{{\texttt http://heasarc.gsfc.nasa.gov/docs/tools.html}} linearly scaled for the pulsar distance (Theureau et al.\ 2010). 
We obtain a 3$\sigma$ upper limit of $F_{\rm X}=5 \times 10^{-14}$ erg cm$^{-2}$ s$^{-1}$ on the unabsorbed flux in the 0.3--10 keV energy band. This value is a factor of 20 deeper than the previous upper limit of $F_{\rm X}=9 \times 10^{-13}$ erg cm$^{-2}$ s$^{-1}$, obtained from {\em Suzaku} observations (Abdo et al.\ 2013) under the same assumptions as above on the source spectrum and absorption.

\section{Discussion}

\subsection{Pulsar distance and extinction}

Both the distance and extinction to these pulsars are not precisely known and an assessment of the estimated values is in order before computing the limits on the pulsar optical luminosities derived from the limiting magnitude of our GTC observations.

For the two radio-loud pulsars (PSR\, J0248+6021 and J0631+1036) no radio parallax measurement has been obtained yet.  Furthermore, it is thought that the distance obtained from the  dispersion measure (DM), inferred from the NE2001 model of the Galactic free electron density along the line of sight  (Cordes \& Lazio 2002), largely overestimates the actual value (Table \ref{dist}). For  PSR\, J0248+6021,  the high DM (370$\pm$1 pc cm$^{-3}$) suggests that the pulsar is within the giant HII region W5 in the Perseus arm, hence its distance must be smaller than that of the far edge of the arm, which is $3.6$ kpc (Reid et al.\ 2009).  However,  Theureau et al.\ (2011) suggest that the pulsar is at the same distance as the open cluster IC\, 1848, which is also within W5. This would imply a distance of 2.0$\pm$0.2 kpc, much lower than the minimum DM-base value D$_{\rm NE}=43.5$ kpc. A distance smaller than inferred from the DM is also suggested by the significant pulsar proper motion (Tab \ref{psr}; Theureau et al.\ 2011).  
For PSR\, J0631+1036, it is speculated that the pulsar is background to the  3-Mon star forming region and, possibly, is within the dark cloud LDN\, 1605 (Zepka et al.\ 1996), which would  substantially contribute to the relatively large DM measured along the line of sight to the pulsar.   Accounting for this contribution would imply a downward revision of the pulsar distance.  In this way, Zepka et al.\ (1996) obtained a distance estimate of 1.0$\pm$0.2 kpc,
a factor of two lower than the value based on the DM (D$_{\rm NE}$=3.7$^{+1.3}_{-0.9}$ kpc). For both pulsars, these alternative distance values have been adopted in  the 2PC (Abdo et al.\ 2013). 

For the radio-quiet PSR\, J0633+0632,  the 2PC gives a maximum distance of 8.7 kpc, determined from the maximum DM along the line of sight and the NE2001 model.  A direct distance estimate to the pulsar can be obtained from the hydrogen column density $N_{\rm H}$ that best fits the \chan\ X-ray spectrum (Ray et al.\ 2011). A fit with a power-law (PL) plus blackbody (BB) model gives  $N_{\rm H}=0.15^{+0.16}_{-0.10} \times 10^{22}$ cm$^{-2}$  (Ray et al.\ 2011).  An independent fit of the same \chan\ data set and with the same spectral model (PL+BB) yields $N_{\rm H}=0.06^{+0.22}_{-0.06} \times 10^{22}$ cm$^{-2}$ (Abdo et al.\ 2013).  Recently, a further re-analysis of the same \chan\ data set  was also carried out by Danilenko et al.\ (2015), who gives $N_{\rm H}=0.24^{+0.18}_{-0.14} \times 10^{22}$ cm$^{-2}$  for a PL+BB model.  Although the $N_{\rm H}$ value of Abdo et al.\ (2013) is somewhat smaller than those of Ray et al.\ (2011) and Danilenko et al.\ (2015), all values are consistent with each other within the statistical uncertainties. By using, e.g. the linear correlation between $N_{\rm H}$ and distance computed by He et al.\ (2013) we obtain a distance of $\approx$ 3 kpc, where the scatter in the $N_{\rm H}$--distance plane 
and the formal uncertainty on the different $N_{\rm H}$ measurements, result in an estimated uncertainty of $\approx$ 1 kpc.  Recently, by assuming that the unknown pulsar proper motion is aligned with the major axis of the X-ray PWN, Danilenko et al.\ (2015) suggested that PSR\, J0633+0632 might have been born in the Rosette Nebula and inferred a tentative pulsar distance as small as 1.2--1.8 kpc. The association, however, has still to be confirmed through a future pulsar proper motion measurement.

For both PSR\, J0248+6021 and PSR\, J0631+1036, which are undetected in the X rays, we used the $N_{\rm H}$ inferred from the DM according to the linear correlation computed by He et al.\ (2013). We note that using three-dimensional Galactic dust maps (e.g., Green et al.\ 2015)  would give reddening values that are obviously distance dependent. Since the distance to these two pulsars is uncertain, the corresponding reddening value is also uncertain.  Furthermore, the Green et al.\ (2015) maps have a resolution of 3\farcm4--13\farcm7, comparable to the field of view of our images (or larger), and might not be representative of the actual reddening towards a specific line of sight 
in the presence of large nebular structures in the field. This is, indeed, the case for both the PSR\, J0248+6021 and PSR\, J0613+1036 fields, with the giant HII region W5 and the dark cloud LDN 1605, respectively.

For the DM towards PSR\, J0248+6021 (370 pc cm$^{-3}$) the correlation yields $N_{\rm H}= 1.11^{+0.48}_{-0.33} \times 10^{22}$ cm$^{-2}$, where the errors are associated with the 90\% confidence interval of the $N_{\rm H}$-DM fit. For PSR\, J0631+1036 the fit yields $N_{\rm H}= 0.37^{+0.16}_{-0.11} \times 10^{22}$ cm$^{-2}$ for a DM of 125.4 pc cm$^{-3}$.
Both values are fully consistent with the qualitative $N_{\rm H}$ estimates obtained from the Galactic value for the pulsar direction, scaled for the distance (Abdo et al.\ 2013). 
From the $N_{\rm H}$, we then estimated the interstellar reddening according to the relation of Predehl \& Schmitt (1995) and obtained $E(B-V) = 1.99^{+0.86}_{-0.39}$ and $E(B-V)=0.66^{+0.28}_{-0.19}$ for PSR\, J0248+6021 and J0631+1036, respectively. %
For PSR\, J0633+0632, the only one of them that has been detected in the X rays, the $N_{\rm H}$ is derived from the spectral fits to the X-ray spectrum (see above).  For instance, assuming the $N_{\rm H}$ estimates obtained  from the X-ray analysis of  Ray et al.\ (2011) and Abdo et al.\ (2013) yields an interstellar reddening $E(B-V)$ of  $0.27^{+0.29}_{-0.18}$ and $0.11^{+0.40}_{-0.11}$, respectively, following Predehl \& Schmitt (1995). Similarly, the  $N_{\rm H}$ obtained from the most recent spectral re-analysis of Danilenko et al.\ (2015) yields an $E(B-V) = 0.43^{+0.32}_{-0.33}$.

The inferred values of the interstellar reddening along the line of sight for the three pulsars are summarised in Table \ref{dist}. For all of them, the reddening is significant, consistently with their estimated distance and low height above the Galactic plane.

\begin{table}
\begin{center}
\caption{Pulsar dispersion measure  (DM) and distance (D$_{\rm NE}$), inferred from the NE2001 model of the Galactic free electron density along the line of sight  (Cordes \& Lazio 2002), of the radio-loud pulsars listed  Table \ref{psr}.  The distance assumed in the Second Catalogue of {\em Fermi} Gamma-ray Pulsars (2PC; Abdo et al.\ 2013) is also given (D$_{\rm 2PC}$). The last column gives the reddening $E(B-V)$ in the pulsar direction estimated from the $N_{\rm H}$ values, obtained either directly from the fit to the X-ray spectrum (PSR\, J0633+0632; Abdo et al.\ 2013) or inferred from the value of the DM (He et al.\ 2013), using the relation of Predehl \& Schmitt (1995).
}
\label{dist}
\begin{tabular}{lcccc} \hline
Pulsar 				 &  DM & D$_{\rm NE}$ & D$_{\rm 2PC}$ & $E(B-V)$ \\ 
                                           &   (pc cm$^{-3}$) & (kpc)  & (kpc)  \\ \hline
J0248+6021   	 & 370$\pm$1 &$>$43.5	   &  2.0$\pm$0.2$^1$ &  1.99$^{+0.86}_{-0.39}$\\
J0631+1036          & 125.4$\pm$0.9         &  3.7 $^{+1.3}_{-0.9}$ & 1.0$\pm$0.2$^2$                     &  0.66$^{+0.28}_{-0.19}$         \\
J0633+0632           &       -      &          -                 &  $<$8.7$^3$&     $0.11^{+0.40}_{-0.11}$     \\ \hline
\end{tabular}
\end{center}
$^1$ Theureau et al.\ (2011); $^2$ Zepka et al.\ (1996); $^3$ estimated from the maximum DM value along the line of sight from the NE2001 model. For PSR\, J0633+0632, independent $N_{\rm H}$ estimates (Ray et al.\ 2011; Danilenko et al.\ 2015) yield $E(B-V)$ values of $0.27^{+0.29}_{-0.18}$ and 
0.43$^{+0.32}_{-0.33}$.
\end{table}

\subsection{Pulsar optical and X-ray luminosities}

We computed the extinction-corrected upper limits on the optical fluxes of the three pulsars in the $g'$ band.  We accounted for the uncertainties on the interstellar reddening (Table \ref{dist}) by assuming the most conservative estimates. From the reddening values, we computed the interstellar extinction in the different filters using the extinction coefficients of Fitzpatrick (1999).  The extinction-corrected flux upper limits are $F_{\rm opt} \sim 4.36 \times 10^{-13}$ erg cm$^{-2}$ s$^{-1}$,  $1.90 \times 10^{-15}$ erg cm$^{-2}$ s$^{-1}$, and  $4 \times 10^{-16}$  erg cm$^{-2}$ s$^{-1}$, for  PSR\, J0248+6021, J0631+1036, and J0633+0632, respectively.  Of course, in the case of PSR\, J0248+6021 which is affected by a much larger interstellar extinction (about 9 magnitudes in the $g'$ band)  the value of the extinction corrected flux is well above that obtained for the other two pulsars. Therefore, the limits on the derived quantities (e.g. optical luminosity and flux ratios) are much less constraining.  For both PSR\, J0248+6021  and PSR\, J0631+1036 we assumed the same distances as in the 2PC (D$_{\rm 2PC}$; see Table \ref{dist}), whereas for PSR\, J0633+0632 we assumed a distance of 3 kpc (Sectn. 4.1). From these values, we computed the corresponding optical luminosity upper limits as  $L_{\rm opt} \sim 2.09 \times 10^{32}$ $d_{\rm 2}^{2}$ erg s$^{-1}$,  $2.28 \times 10^{29}$ $d_{\rm 1}^{2}$ erg s$^{-1}$  and $4.3 \times 10^{29}$ $d_{\rm 3}^{3}$ erg s$^{-1}$, for  PSR\, J0248+6021, J0631+1036, and J0633+0632, respectively, where $d_{\rm 1}$, $d_{\rm 2}$, and $d_{\rm 3}$ are their distances in units of 1, 2, and 3 kpc. We remind that the assumed distance values for the two radio-loud pulsars PSR\, J0248+6021 and PSR\, J0631+1036 have been obtained from indirect estimates (see Theureau et al.\ 2010 and Zepka et al.\ 1996).  As such, they might be affected by uncertainties larger than those associated with the assumed values (Table \ref{dist}), which might imply correspondingly higher luminosities for these two pulsars. In particular, we cannot firmly rule out that PSR\, J0248+6021 and PSR\, J0631+1036 are at distances as high as 3.6 and 5 kpc, respectively,  the former corresponding to the far edge of the Perseus arm and the latter to the maximum value of the DM distance D$_{\rm NE}$ (Table \ref{dist}).  Direct  distance measurements will be crucial to better constrain their luminosities. 
We compared the optical luminosity upper limits with the pulsar spin-down energy $\dot{E}$. Since these pulsars are younger than 0.1 Myr (Table \ref{psr}), we assume that their optical luminosity is dominated by non-thermal emission, as usually observed in pulsars of comparable age (see e.g., Mignani 2011), and is powered by the spin-down energy.  We obtained  $L_{\rm opt}/\dot{E} \la 9.93 \times 10^{-4} d_{\rm 2}^2$,  $\la 1.34 \times 10^{-6} d_{\rm 1}^2$, and $\la 3.58 \times 10^{-6} d_{\rm 3}^2$, for  PSR\, J0248+6021, J0631+1036, and J0633+0632, respectively.

Both the optical luminosity and the optical emission efficiency, $L_{\rm opt}/\dot{E}$, of rotation-powered pulsars are strongly correlated with the spin-down age (see, e.g. Fig.\, 4 in Kirichenko et al.\ 2015).
As expected, owing to their lower spin-down energy ($\dot{E} \approx 10^{35}$ erg s$^{-1}$), both PSR\, J0631+1036 and J0633+0632 are fainter in the optical than the young ($\tau \la 5$ kyr) and more energetic ($\dot{E} \approx 10^{37}$--$10^{38}$ erg s$^{-1}$) $\gamma$-ray pulsars Crab, PSR\, B1509$-$58,  and PSR\, J0205+6449, which have optical luminosities $L_{\rm opt} \approx  10^{30}$--$10^{33}$ erg s$^{-1}$ (e.g. Moran et al.\ 2013). The brightest of them is now PSR\, B0540$-$69 ($\sim$2 kyr),  only recently detected as a $\gamma$-ray pulsar  (Ackermann et al.\ 2015), which has an optical luminosity $L_{\rm opt} \sim 2.6\times10^{33}$ erg s$^{-1}$.  Only  PSR\, J0248+6021 could be, in principle,   as luminous as the very young pulsars, owing to the less deep constraints on its optical luminosity.  Depending on their actual distances, the three pulsars discussed in this work might have a lower efficiency in converting their spin down power into optical radiation  than the Crab and  PSR\, B0540$-$69 ($L_{\rm opt}/\dot{E} \approx 10^{-5}$). This  would indicate that both the optical luminosity and emission efficiency $L_{\rm opt}/\dot{E} $ rapidly decrease for pulsars older than $\sim$ 5 kyr,  but younger than $\sim 0.1$ Myr, an hypothesis initially suggested by the characteristics of the Vela pulsar  (see discussion in Mignani et al.\ 1999), which has an optical luminosity $L_{\rm opt} \sim 1.35 \times 10^{28}$ erg s$^{-1}$ and an $L_{\rm opt}/\dot{E} \sim 1.9 \times 10^{-9}$ (e.g. Moran et al.\ 2013). This hypothesis seems to be supported by the recent detection of the candidate optical counterpart to PSR\, J0205+6449 ($\sim 5.4$ kyr), which has an an optical luminosity $L_{\rm opt} \sim 1.15 \times 10^{30}$ erg s$^{-1}$ and an $L_{\rm opt}/\dot{E}  \sim 4.2 \times 10^{-8}$, ideally linking the Crab-like pulsars to Vela. The corresponding upper limits on $L_{\rm opt}$ and  $L_{\rm opt}/\dot{E} $ for PSR\, J0248+6021, PSR\, J0631+1036, and PSR\, J0633+0632, however, are all above the corresponding values for the Vela pulsar, so that we cannot rule out that some of them are more luminous and convert their spin-down energy into optical radiation more efficiently, unless they are at  significantly lower distance than estimated.  The larger distance of these pulsars and/or the larger interstellar extinction towards the line of sight with respect to Vela (Table \ref{dist}) are the major obstacle to obtain tighter constraints on their optical luminosity and strengthen their similarity to the Vela pulsar.

\begin{figure}
\centering
{\includegraphics[width=8.5cm,angle=0,clip=]{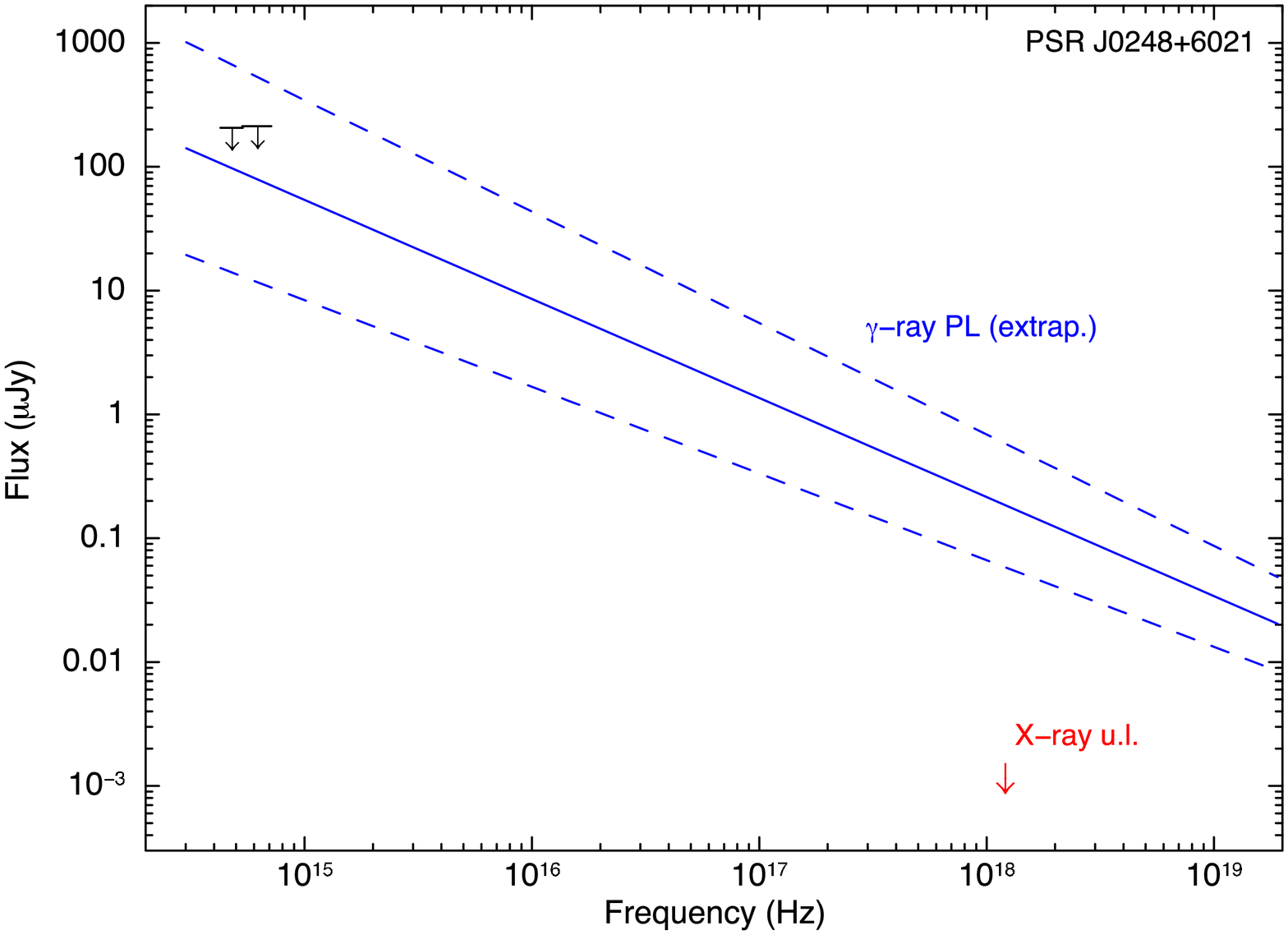}}
{\includegraphics[width=8.5cm,angle=0,clip=]{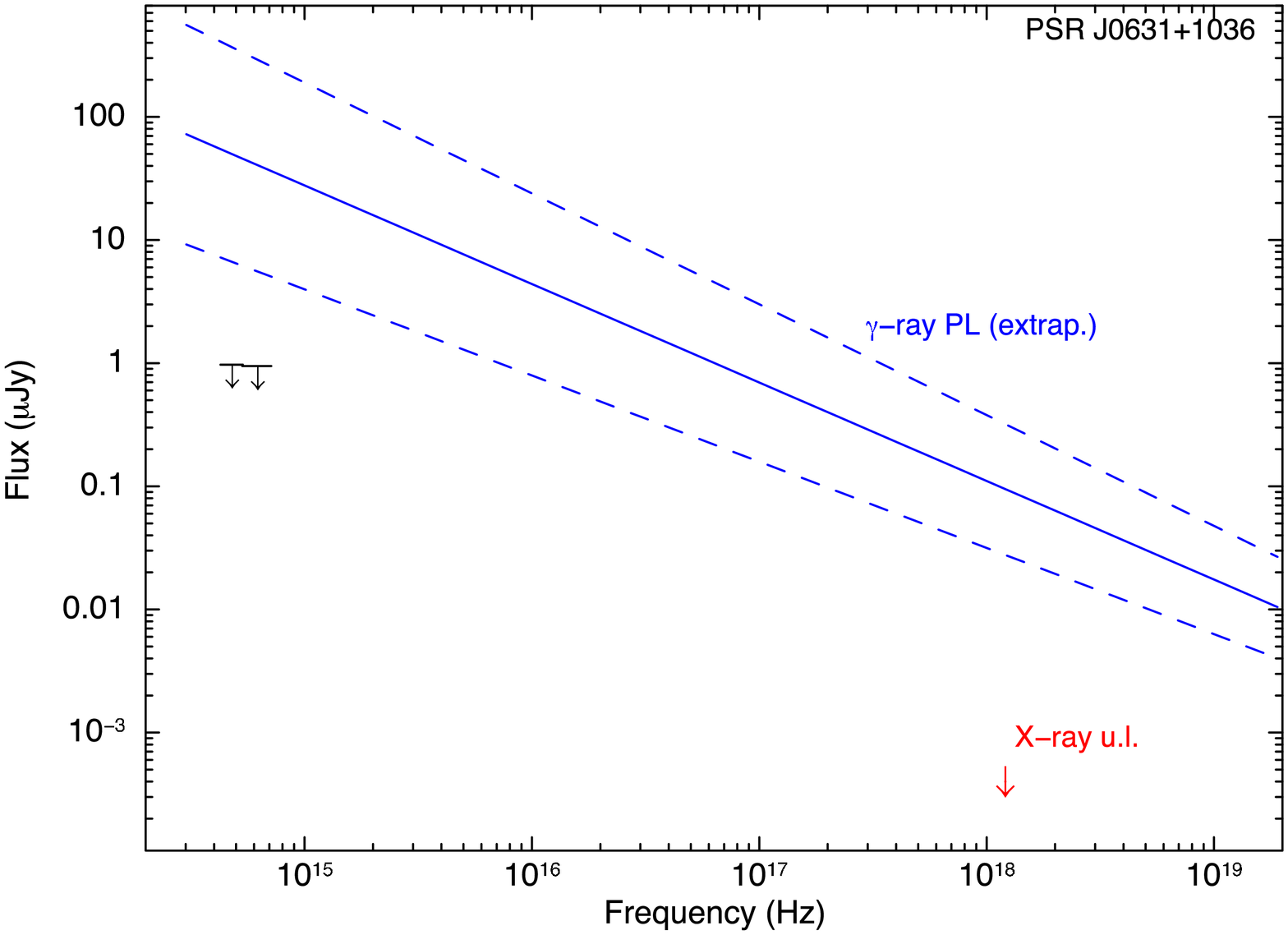}}
{\includegraphics[width=8.5cm,angle=0,clip=]{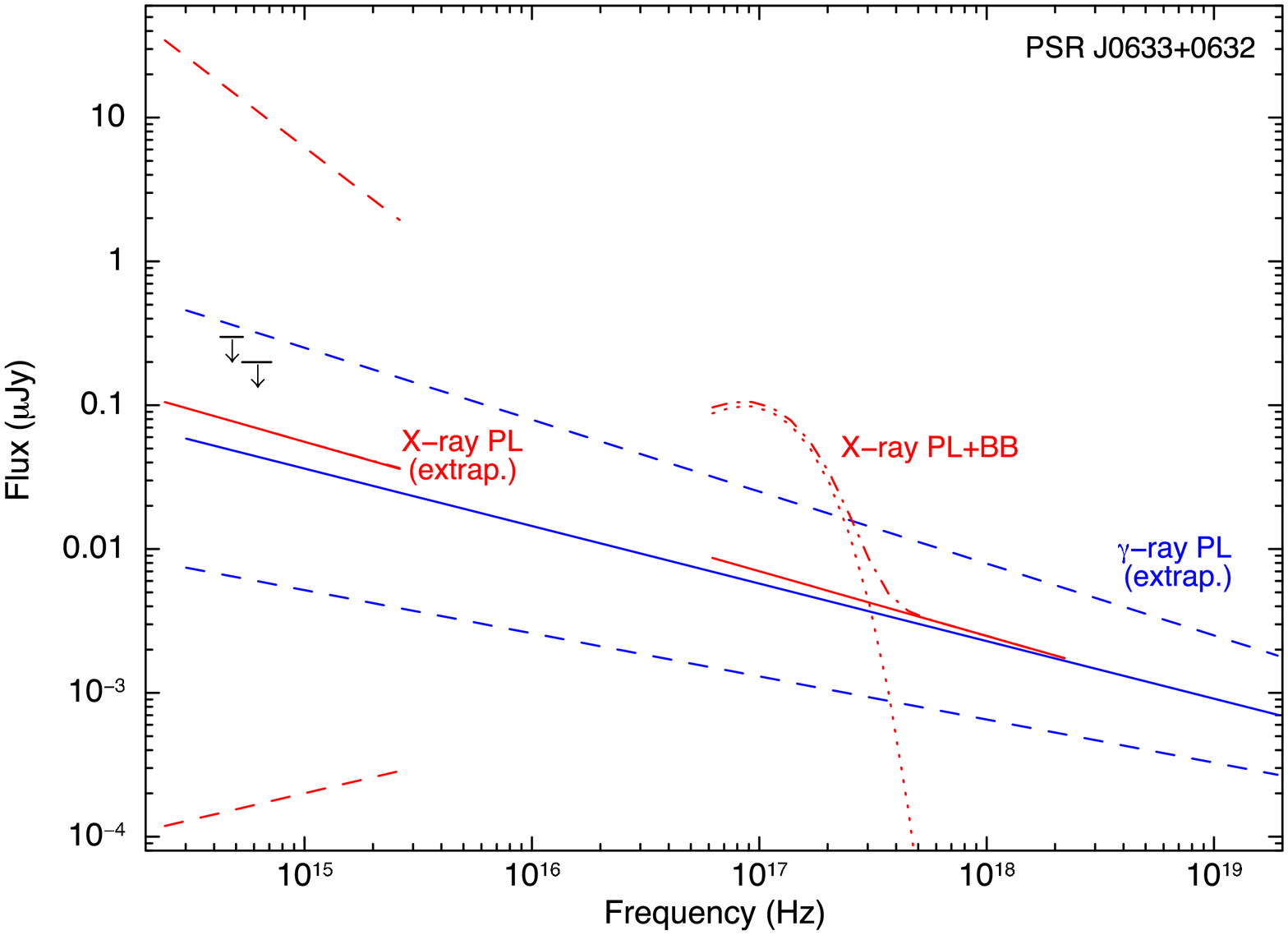}}
\caption{\label{psr-sed} 
Extinction-corrected optical flux upper limits (black) of PSR\, J0248+6021, PSR\, J0631+1036, and PSR\, J0633+0632 (top to bottom) compared with the extrapolation in the optical domain of the $\gamma$-ray PL spectrum (in blue) and of the the X-ray PL component (in red). In both cases, the best-fit PL is represented by the solid line and the dotted lines represent the $1 \sigma$ uncertainty. For PSR\, J0633+32, the dotted line represents the BB component and the dot-dashed line the total BB+PL spectrum.  The red arrows in the upper and middle panels correspond to the unabsorbed X-ray flux upper limit at 5 keV for an assumed PL X-ray spectrum with photon index $\Gamma_{\rm X}=2$. }
\end{figure}

Our upper limit on the unabsorbed X-ray flux of PSR\, J0248+6021 in the 0.3--10 keV energy band ($F_{\rm X} \la 5 \times 10^{-14}$ erg cm$^{-2}$ s$^{-1}$; Sectn.\ 3.2) corresponds to a non-thermal X-ray luminosity $L_{\rm X} \la 5.9 \times 10^{30} d_{\rm 2}^{2}$ erg  s$^{-1}$. 
This implies an  X-ray emission efficiency $L_{\rm X}/\dot{E} \la 2.8 \times 10^{-5} d_{\rm 2}^{2}$, which suggests that PSR\, J0248+6021 converts a significant lower fraction of its spin-down energy in X rays with respect to the bulk of the rotation-powered pulsars (see, discussion in Becker 2009). For comparison, in the case of PSR\, J0631+1036 the  limit on its non-thermal 0.3--10 keV X-ray flux ($F_{\rm X} \la 2.3 \times 10^{-14}$ erg cm$^{-2}$ s$^{-1}$; Abdo et al.\ 2013), also computed  for a $\Gamma_{\rm X}=2$ and a linear scaling of the Galactic $N_{\rm H}$, yields an X-ray luminosity $L_{\rm X} \la 0.68 \times 10^{30} d_{\rm 1}^{2}$ erg  s$^{-1}$ and an $L_{\rm X}/\dot{E} \la 4.0 \times 10^{-6} d_{\rm 1}^{2}$ erg  s$^{-1}$. Both their low X-ray luminosity and X-ray emission efficiency help to explain the non-detection of both pulsars in the available X-ray observations. For instance,  PSR\, J0633+0632 has an  X-ray luminosity $L_{\rm X} =(1.7 \pm 0.13) \times 10^{31} d_{\rm 3}^{2}$ erg  s$^{-1}$, computed  from its unabsorbed non-thermal 0.3--10 keV X-ray flux $F_{\rm X}=(6.3\pm0.5) \times 10^{-14}$ erg cm$^{-2}$ s$^{-1}$ (Abdo et al.\ 2013),
which is higher than the other two pulsars. Owing to a comparable spin-down energy, this also yields an  higher X-ray emission efficiency $L_{\rm X}/\dot{E} \sim (1.4\pm0.1) \times 10^{-4} d_{\rm 3}^{2}$. 

For all the three pulsars discussed in this work, the luminosity and luminosity--to--spin-down energy ratios in both the optical and X rays are summarised in the first part of Table \ref{mw}.

\subsection{Pulsar multi-wavelength spectra}

Outside the radio band, both PSR\, J0248+6021 and PSR\, J0631+1036 have been detected only in $\gamma$-rays. In both cases, we compared the extinction-corrected optical flux upper limits with the extrapolation in the optical domain of their  $\gamma$-ray spectra. These are described by a power law with exponential cut-off, with photon index $\Gamma_{\gamma}=1.8 \pm 0.01$  and cut-off energy $E_c= 1.6 \pm  0.03$ GeV, for PSR\, J0248+6021, and $\Gamma_{\gamma}=1.8 \pm 0.01$  and cut-off energy $E_c= 6 \pm  1$ GeV, for PSR\, J0631+1036 (Abdo et al.\ 2013). The corresponding energy fluxes $F_{\gamma}$ in the 0.1 to 100 GeV energy band are $(5.2\pm0.4)\times 10^{-11}$ erg cm$^{-2}$ s$^{-1}$ and $(4.7\pm0.3)\times 10^{-11}$ erg cm$^{-2}$ s$^{-1}$, respectively.  The spectral energy distributions (SEDs)  of PSR\, J0248+6021 and PSR\, J0631+1036 are shown in Fig. \ref{psr-sed} (top and mid panel, respectively). For PSR\, J0248+6021, the optical flux upper limits are within the uncertainties of the extrapolation of the best-fit $\gamma$-ray PL spectrum. We remind, however, that owing to the large interstellar extinction correction, the optical flux upper limits are not very constraining. Therefore, if one considers only the $\gamma$-ray and optical flux measurements nothing could be said about the presence, or absence, of a spectral break in the pulsar non-thermal emission between the optical and the $\gamma$-ray regions.
For PSR\, J0631+1036, the picture is different, with the optical flux upper limits lying well below the uncertainty on the extrapolation of the $\gamma$-ray PL. This obviously indicates a spectral break 
between the optical and the $\gamma$ rays.
For both pulsars, we constrained the SED in this energy interval from their non-detection in the X rays. Assuming for both of them the same X-ray spectrum as assumed in Sectn.\, 3.2 and 4.2, i.e. a PL with photon index $\Gamma_{\rm X}=2$, would imply an upper limit on the unabsorbed X-ray flux at 5 keV of $1.2\times10^{-3}$ $\mu$Jy for PSR\, J0248+6021 and of $0.53\times10^{-3}$ $\mu$Jy for PSR\, J0631+1036 (red arrows in Fig.\ref{psr-sed}). In both cases, these limits are well below the extrapolation of the $\gamma$-ray PL, which indicates the presence of at least a spectral break between the optical and the $\gamma$-rays. We note that assuming a different PL photon index does not alter this scenario. 
Detecting both PSR\, J0248+6021 and PSR\, J0631+1036  in the X rays would be crucial 
constrain the energy at which the break occurs. 

PSR\, J0633+0632 has been detected both in the $\gamma$ and X rays.  It has a $\gamma$-ray  photon index similar to the other two pulsars, $\Gamma_{\gamma}=1.4 \pm 0.1$, and a cut-off energy $E_c= 2.7 \pm  0.3$ GeV. Its energy flux $F_{\gamma} = (9.4\pm0.5)\times 10^{-11}$ erg cm$^{-2}$ s$^{-1}$ makes it the brightest of these three pulsars in  $\gamma$ rays. 
 For the X-ray spectrum, a fit to the \chan\ data with a PL+BB spectral model yields a photon index $\Gamma_{\rm X} = 1.45^{+0.76}_{-0.82}$ and  a temperature $k T = 0.126^{+0.024}_{-0.033}$ keV (Abdo et al.\ 2013), corresponding to an unabsorbed total (BB+PL) X-ray flux $F_{\rm X}=(1.71 \pm 0.14) \times 10^{-13}$ erg cm$^{-2}$ s$^{-1}$ in the 0.3--10 keV energy band.
The original X-ray analysis in Ray et al.\ (2011) yielded $\Gamma_{\rm X} = 1.5\pm0.6$ and   $k T = 0.11^{+0.03}_{-0.02}$ keV.  These spectral parameters are also very similar to those obtained by Danilenko et al.\ (2015) for the same spectral model,  $\Gamma_{\rm X} = 1.6\pm0.6$ and $k T = 0.105^{+0.023}_{-0.018}$ keV.
The SED of PSR\, J0633+0632 (Fig. \ref{psr-sed} bottom panel) again shows, that the optical flux upper limit are within the extrapolation of the $\gamma$-ray PL. The comparison with the X-ray PL extrapolation is, however, extremely uncertain owing to the 
large uncertainties on the photon index $\Gamma_{\rm X}$. A more robust X-ray detection of PSR\, J0633+0632 would be needed to decrease the uncertainties on the PL spectrum and determine whether this is, indeed, consistent with the extrapolation of the $\gamma$-ray PL.  Were this the case, we would have a rare example of a pulsar where a single PL can (possibly) describe its entire non-thermal emission.

\begin{table*}
\begin{center}
\caption{Multi-wavelength properties of the pulsars discussed in this work. Columns  two and three report the upper limits on the optical luminosity derived from the GTC observations (Sectn.\ 3.1) and the upper limit on the X-ray luminosity of PSR\, J0248+6021 derived from the \chan\ observations (Sectn.\ 3.2).  The X-ray luminosity value and upper limit for PSR\, J0633+0632 and  PSR\, J0631+1036, respectively, have been computed from their unabsorbed non-thermal X-ray flux $F_{\rm X}$ as given in the 2PC (Abdo et al.\  2013). Luminosity values are scaled for the assumed distance (Sectn.\ 4.1). The $F_{\rm opt}/F_{\gamma}$  and  $F_{\gamma}/F_{\rm X}$ ratios have been computed from the pulsar $\gamma$-ray flux $F_{\gamma}$ in the 2PC.  All limits conservatively accounts for the statistical uncertainties on the measured flux values. The hyphen marks the cases where the flux ratio is unconstrained. }
\label{mw}
\begin{tabular}{lccccccc} \hline
Pulsar 			&	$L_{\rm opt}$ & $L_{\rm X}$ & $L_{\rm opt}/\dot{E}$ & $L_{\rm X}/\dot{E}$  &   $F_{\rm opt}/F_{\rm X}$ & $F_{\rm opt}/F_{\gamma}$  & $F_{\gamma}/F_{\rm X}$ \\  
                                  &      ($10^{30}$ erg  s$^{-1}$)                       & ($10^{30}$ erg  s$^{-1}$)  & & & & & \\ \hline                                  
J0248+6021   	         &  $\la 209d_{\rm 2}^{2}$    & $\la 5.9 d_{\rm 2}^{2}$ &  $\la 9.93 \times 10^{-4} d_{\rm 2}^2$      & $\la 2.8 \times 10^{-5} d_{\rm 2}^{2}$ & - & $\la 9.08\times 10^{-3}$& $\ga 960$ \\
J0631+1036              &  $\la 0.228     d_{\rm 1}^{2}$ & $\la 0.68 d_{\rm 1}^{2}$ &  $\la 1.34 \times 10^{-6} d_{\rm 1}^2$    & $\la 4.0 \times 10^{-6} d_{\rm 1}^{2}$  &- &$\la 4.05 \times 10^{-5}$& $\ga 1900$ \\   
J0633+0632              &   $ \la 0.43  d_{\rm 3}^{2}$   &$(17 \pm 1.3)  d_{\rm 3}^{2}$ &  $\la 3.58 \times 10^{-6} d_{\rm 3}^2$   & $(1.4\pm 0.1) \times 10^{-4} d_{\rm 3}^{2}$ & $\la 6.9 \times 10^{-3}$ &$\la 4.5 \times 10^{-6}$& 1510$\pm$170 \\ \hline
\end{tabular}
\end{center}
\end{table*}

\subsection{Pulsar multi-wavelength emission}

We characterised the multi-wavelength emission properties of these three pulsars from the ratios between their extinction-corrected optical and X-ray fluxes $F_{\rm opt}$ and $F_{\rm X}$ and between these and their $\gamma$-ray flux $F_{\gamma}$ (Sectn.\ 4.3). We note that such ratios are equivalent to luminosity ratios, with the advantage that they are obviously independent on the uncertainty on the pulsar  distance. 

We first used the upper limits on the extinction-corrected pulsar optical fluxes ($g'$ band) to constrain the ratio with their $\gamma$-ray flux $F_{\gamma}$.   We obtained $F_{\rm opt}/F_{\gamma} \la 9.08\times 10^{-3}$, $\la 4.05 \times 10^{-5}$, and $\la 4.5 \times 10^{-6}$ for PSR\, J0248+6021, PSR\, J0631+1036, and PSR\, J0633+0632, respectively.

Only for PSR\, J0633+0632 we can constrain the ratio between the pulsar unabsorbed optical and X-ray flux, which is $F_{\rm opt}/F_{\rm X}  \la 6.9 \times 10^{-3}$. 
We compared the unabsorbed optical--to--X-ray and optical--to--$\gamma$-ray flux (luminosity) ratios with the corresponding values for all the $\gamma$-ray pulsars detected in the optical (see, e.g. Moran et al.\ 2013 for a summary), the group now including PSR\, B0540$-$69. For this pulsar, the $\gamma$-ray flux above 0.1 GeV is $F_{\gamma} = (2.6\pm0.3) \times 10^{-11}$ erg cm$^{-2}$ s$^{-1}$ (Ackermann et al.\ 2015).  The unabsorbed optical flux, obtained from {\em HST} observations (Mignani et al.\ 2010),  is $F_{\rm opt} = 8.79 \times 10^{-15}$ erg cm$^{-2}$ s$^{-1}$, whereas the unabsorbed X-ray flux in the 0.3--10 keV energy band is $F_{\rm X}=1.05 \times 10^{-11}$ erg cm$^{-2}$ s$^{-1}$, as computed from the \chan\ observations (Kaaret et al.\ 2001).
This gives unabsorbed flux ratios of $F_{\rm opt}/F_{\gamma} \sim 3.4 \times 10^{-4}$ and $F_{\rm opt}/F_{\rm X} \sim 8.4 \times 10^{-4}$. 
We note that for the younger pulsars (Crab, PSR\, B0540$-$69, PSR\, B1509$-$58, PSR\, J0205+6449) the optical and X-ray emission are both non-thermal, whereas for the Vela pulsar the optical emission is non-thermal, while thermal emission from the neutron star surface partially accounts for the X-ray emission (e.g., Mignani 2011; Becker 2009). Therefore, one has to account for the non-thermal X-ray emission only, which is about 1/4 of the total (Marelli et al.\ 2011), and the unabsorbed optical--to--X-ray flux ratio for Vela is, then, $F_{\rm opt}/F_{\rm X} \sim 2.1 \times 10^{-4}$. For the middle-aged pulsars (PSR\, B0656+14, Geminga, PSR\, B1055$-$52) the optical and X-ray emission are both produced by the combination of thermal and non-thermal  processes  (Mignani 2011; Becker 2009) and we assume the unabsorbed total fluxes at both energies as a reference, as in Moran et al.\ (2013).

Apart from PSR\, J0248+6021, for which the limit on the optical flux is very conservative, the upper limits on the $F_{\rm opt}/F_{\gamma}$ for both PSR\, J0631+1036 and PSR\, J0633+0632 indicate that in pulsars other than the young, Crab-like, ones  the optical emission becomes lower and lower with respect to the $\gamma$-ray one, with  the $F_{\rm opt}/F_{\gamma}$  decreasing from $\sim 1.4 \times 10^{-5}$ to $\sim  6.4 \times 10^{-8}$ as the spin-down age increases from a few kyrs to about a Myr (see Table 4 of Moran et al.\ 2013).  This is likely related to the fact that efficiency in converting spin-down energy into $\gamma$-ray radiation is larger for middle-aged pulsars than for the young ones (Abdo et al.\ 2013). The upper limits on the $F_{\rm opt}/F_{\gamma}$ ratio for these two pulsars are still above the corresponding value of the Vela pulsar ($\sim 1.6 \times 10^{-7}$), so that we cannot rule out that they are intrinsically brighter in the optical than in $\gamma$ rays with respect to Vela, despite of they larger spin-down age. The $F_{\rm opt}/F_{\rm X}$ ratio for PSR\, J0633+0632 is $\la 6.9 \times 10^{-3}$, which is by far the highest of any other $\gamma$-ray pulsar (let alone all other isolated neutron stars), making this pulsar a very interesting case to study. Much deeper optical observations, challenging for current facilities, would be crucial to determine whether the actual value of the $F_{\rm opt}/F_{\rm X}$ ratio is more in line with that of all the other optical/X-ray emitting $\gamma$-ray pulsars.

\begin{figure}
\centering
{\includegraphics[height=6.5cm,angle=0,clip=]{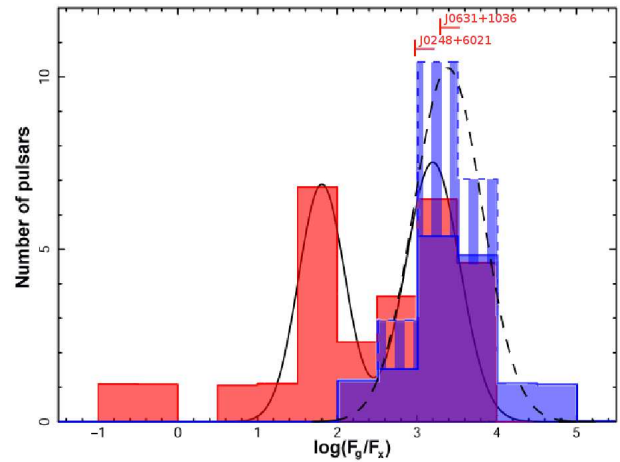}}
\caption{\label{histo} 
Histogram of the $F_{\gamma}/F_{\rm X}$ ratios for all young--to--middle-aged $\gamma$-ray pulsars (adapted from Marelli et al.\ 2015). Radio quiet-pulsars are shown in blue and radio-loud pulsars in red. The dashed and continuous curves are the Gaussian functions best-fitting the peaks in the distribution. The limits for the two radio-loud pulsars  PSR\, J0248+6021  and PSR\, J0631+1036 are indicated. }
\end{figure}

We also compared the ratios between the pulsar $\gamma$ and X-ray fluxes.
Our new \chan\ limit on the unabsorbed X-ray flux of PSR\, J0248+6021  ($F_{\rm X} \la 5 \times 10^{-14}$ erg cm$^{-2}$ s$^{-1}$; Sectn.\ 3.2) raises the lower limit on its $\gamma$--to--X-ray flux ratio $F_{\gamma}/F_{\rm X}$ to $\sim 960$. This value is consistent with the largest $F_{\gamma}/F_{\rm X}$ values in the second peak of the two-peak distribution observed in radio-loud pulsars (see Fig.\ref{histo} and Marelli et al.\ 2015). For comparison, the  unabsorbed X-ray flux upper limit ($F_{\rm X} \la 2.3 \times 10^{-14}$ erg cm$^{-2}$ s$^{-1}$; Abdo et al.\ 2013) on the other radio-loud pulsar, PSR\, J0631+1036 yields $F_{\gamma}/F_{\rm X}\ga 1900$. Both limits are also consistent with the peak in the $F_{\gamma}/F_{\rm X}$ distribution of radio-quiet pulsars (Fig.\ref{histo}). For instance, with an unabsorbed non-thermal X-ray flux $F_{\rm X}=(6.3\pm0.5) \times 10^{-14}$ erg cm$^{-2}$ s$^{-1}$ (Abdo et al.\ 2013), the radio-quiet PSR\, J0633+0632 has a measured $F_{\gamma}/F_{\rm X}=1510\pm170$. The lower limits on the  $F_{\gamma}/F_{\rm X}$ for both PSR\, J0248+6021  and PSR\, J0631+1036 confirm that, at variance with others, some radio-loud pulsars are more similar to the radio-quiet ones in their high-energy behaviour. This is shown by the double-peaked  $F_{\gamma}/F_{\rm X}$ distribution for radio-loud pulsars, with the second peak overlapping the peak of the corresponding distribution for the radio-quiet ones (Fig.\ref{histo}).  This similarity in the $F_{\gamma}/F_{\rm X}$ ratios does not reflect a similarity in the pulsar characteristics (e.g., spin period, age, magnetic field) but it is possibly related to a similar geometry of the X and $\gamma$-ray emission regions  (see discussion in Marelli et al.\ 2015). 

The multi-wavelength flux ratios for these three pulsars are summarised in the second part of Table \ref{mw}.

\subsection{Comparison with the Vela-like pulsars}

It is now interesting to compare the properties of PSR\, J0248+6021, J0631+1036, and J0633+0632 with those of the slightly younger, Vela-like $\gamma$-ray pulsars for which at least deep optical upper limits with 10m-class telescopes have been obtained after the launch of {\em Fermi} (see Table \ref{vela} for a compilation). These are: PSR\, J0007+7303, J1028$-$5819, B1046$-$58, J1357$-$6429, and J2021+3651 (Mignani et al.\ 2013; 2012, Razzano et al.\ 2013; Mignani et al.\ 2012; Kirichenko et al.\ 2015). To this sample we must add PSR\, B1706$-$44 that was observed during the commissioning of the VLT  (Mignani et al.\ 1999) after its original detection as a $\gamma$-ray pulsar by the {\em CGRO} (Thompson et al.\ 1992) but not re-observed with any facility ever since. Of these pulsars, only PSR\, J0007+7303 and J0633+0632 are radio quiet. 

We compared the optical emission properties of the three pulsars discussed in this work with those of the Vela-like $\gamma$-ray pulsars mentioned above. The limits on the optical luminosity $L_{\rm opt}$ and on $L_{\rm opt}/\dot{E}$ for PSR\, 
J0631+1036 and PSR\, J0633+0632 are comparable, for the assumed values of distance and interstellar extinction, to those of the Vela-like pulsars, 
which are in the range 
$\sim0.04$--$0.92 \times 10^{30}$ erg s$^{-1}$ and $9.4 \times 10^{-8}$--$1.1 \times 10^{-6}$, 
respectively. This confirms, on a broader sample, that  pulsars in the age range 10--100 kyrs  are fainter and less efficient in the optical than the Crab-like ones (see also, Danilenko et al.\ 2013; Kirichenko et al.\ 2015). 
The limits on the  $F_{\rm opt}/F_{\rm X}$ and $F_{\rm opt}/F_{\gamma}$ ratios for both PSR\, J0631+1036 and PSR\, J0633+0632 are also quite similar, indicating that the optical emission always tends to be less than $\approx 10^{-3}$ and $\approx 10^{-6}$ of the X and $\gamma$-ray ones, respectively. Only for PSR\, J2021+3651 the $F_{\rm opt}/F_{\rm X}$  could be as high as a few $10^{-2}$ (Kirichenko et al.\ 2015), whereas the  limits on the $F_{\rm opt}/F_{\gamma}$ are more in line with those of the other pulsars.

\begin{table*}
\begin{center}
\caption{Characteristics and optical properties of the Vela pulsar (in bold) and all Vela-like pulsars observed with 10m-class telescopes. Parameters and units are the same as in Table \ref{psr} and \ref{mw}. The optical luminosity values ($L_{\rm opt}$) are scaled for the pulsar distance assumed in the reference publications (last column), following the same notation  as in Table \ref{mw}. All limits have been computed for the maximum value of the assumed interstellar extinction.}
\label{vela}
\begin{tabular}{lcrrrrllrrl} \hline
Pulsar 			&	P$_{\rm s}$	&    $\dot{P_{\rm s}}$ 	& $\tau$ & B & $\dot{E}$  & $L_{\rm opt}$     &  $L_{\rm opt}/\dot{E}$ &   $F_{\rm opt}/F_{\rm X}$ & $F_{\rm opt}/F_{\gamma}$  & Refs. \\   \hline
J0007+7303		&0.315 & 36.04 & 1.39 &  10.8 & 4.5 & $\la 0.04 d_{1.7}^2$	   & $\la 9.4 \times 10^{-8} d_{1.7}^2$ & $\la 1.6 \times 10^{-3}$  & $\la 5.9  \times 10^{-6}$  & Mignani et al.\ (2013) \\
{\bf B0833$-$45}            & 0.089 & 12.5 & 1.13 & 3.38 & 69.6 & $0.01 d_{0.29}^2$ & $1.9 \times 10^{-9} d_{0.29}^2$  & $5.9 \times 10^{-5}$ & $1.6 \times 10^{-7}$ & Moran et al.\  (2013) \\
J1028$-$5819         & 0.091& 1.61 &  9 & 1.23 & 8.3 & $\la 0.92 d_{3}^2$         & $\la 1.1 \times 10^{-6} d_{3}^2$   &  $\la 9.7 \times 10^{-3}$  & $\la 3.7 \times 10^{-6}$ & Mignani et al.\ (2012) \\
B1046$-$58            & 0.123 & 9.63 & 2.03 &  3.49 & 20 & $\la 0.38 d_{3}^2 $        & $\la 1.9 \times 10^{-7} d_{3}^2$ &  $\la 4.1 \times 10^{-4}$   &  $\la 1.8 \times 10^{-6}$  & Razzano et al.\ (2013) \\
J1357$-$6429        & 0.166 & 36.02 & 0.73 & 7.83 & 31 & $\la 2.16 d_{3}^2$       &  $\la 6.9 \times 10^{-7} d_{3}^2$ &  $\la 4.8 \times 10^{-4}$     & $\la 5.9 \times 10^{-7}$ & Mignani et al.\ (2011) \\
B1706$-$44          &  0.102 & 9.29 & 1.75 & 3.12 & 34 & $\la 0.25 d_{2.3}^2$    &  $\la 7.4 \times 10^{-8} d_{2.3}^2$  & $\la 1.1 \times 10^{-3}$     & $\la 2.9 \times 10^{-7}$ & Mignani et al.\ (1999) \\ 
J2021+3651          &  0.103 & 9.57 & 1.72 & 3.19 & 34 & $\la 0.79 d_{3.5}^2$   &  $\la 2.3 \times 10^{-7} d_{3.5}^2$   & $\la 6.3 \times 10^{-2}$   & $\la 1.4 \times 10^{-6}$ & Kirichenko et al.\ (2015) \\
\hline 
\end{tabular}
\end{center}
\end{table*}

The SEDs of PSR\, J0248+6021, J0631+1036, and J0633+0632 (Sectn.\ 4.3) do not follow an unique template, with the presence of a clear spectral break between the optical and $\gamma$-rays for PSR J0631+1036  and PSR J0248+6021, and with a possible spectral continuity across the optical/X-ray/$\gamma$-ray range for PSR\, J0633+0632.
 This is in line with the behaviour of the slightly younger, Vela-like $\gamma$-ray pulsars, which also seem to be quite different from each other in terms of number of breaks in the multi-wavelength SED. For instance, both PSR\, B1706$-$44 and PSR\, J1028$-$5819 (Mignani et al.\ 2012) feature two breaks in the SED, between the $\gamma$ rays and the X rays and between the X rays and the optical. On the other hand, both PSR\, J0007+7303 (Mignani et al.\ 2013) and PSR\, J2021+3651 (Kirichenko et al.\ 2015) seem to feature only one break, i.e. from the $\gamma$ rays to the X rays, whereas the optical flux upper limits are compatible with the extrapolation of the X-ray PL spectrum. In all cases, the optical flux upper limits lies either above or below the extrapolation of the $\gamma$-ray PL spectrum in the optical domain.  This suggests that, at least in these cases,  the optical and $\gamma$-ray emission are not related. A single PL spectrum running from the optical to the $\gamma$ rays, however, cannot be ruled out for both PSR\, B1046$-$58 (Razzano et al.\ 2013) and PSR\, J1357$-$6429 (Mignani et al.\ 2011).  The multi-wavelength SED of the Vela pulsar (see Fig.\ 7 of Danilenko et al.\ 2011) features a clear spectral break between the X rays and the optical, whereas the extrapolation of the $\gamma$-ray PL  overshoots the X-ray spectrum, indicating the existence of a second break.

Such a difference in the multi-wavelength behaviour of Vela-like $\gamma$-ray pulsars might depend on something other than the pulsar characteristics (e.g., spin period, age, magnetic field), which are similar for  most of them (Table \ref{vela}). One possibility is that it might be related to a difference in the geometry of the optical/X/$\gamma$-ray emission regions in the neutron star magnetosphere, possibly related to the relative inclination between the magnetic and spin-axis,  and/or in the viewing angle. In this case, one would expect a correlation between the observed number of breaks in the non-thermal multi-wavelength SED and the changes in the light curve morphology, such as the relative phase offsets and separation between the peaks.  In the case of the Vela pulsar, the light curve profile indeed changes from the $\gamma$ rays to the X rays and from the X rays to the optical (Abdo et al.\ 2009b), in coincidence with the observed  breaks in the multi-wavelength SED (Danilenko et al.\ 2011). This case is different, however, from that of the younger Crab pulsar for which a clear spectral break is visible only between the X rays and the optical but both the light curve profile and the phase alignment between the two peaks do not vary appreciably between these two energy bands (Abdo et al.\ 2010b).  Unfortunately,  verifying such an hypothetical correlation between spectral breaks and light curve morphologies for the Vela-like $\gamma$-ray pulsars is not straightforward. Both PSR\, J1028$-$5819 and  PSR\, B1046$-$58 have not been yet detected as X-ray pulsars (Mignani et al.\ 2012; Gonzalez et al.\ 2006), whereas both PSR\, J0007+7303 and PSR\, J1357$-$6429  are X-ray pulsars but the X-ray pulsations are thermal in origin (Caraveo et al.\ 2010; Chang et al.\ 2012). Therefore, it is not possible to directly compare their X-ray and $\gamma$-ray light curves.  Finally, for both PSR\, B1706$-$44 and PSR\, J2021+3651 it is difficult to disentangle the contribution of thermal and non-thermal emission to the X-ray light curve (Gotthelf et al.\ 2002; Abdo et al.\ 2009c). Furthermore, none of them has been obviously detected as an optical pulsar yet.  
More (and deeper) multi-wavelength observations are necessary to determine the connection (if any) between spectral breaks and variations in the light curve profiles as a function of energy, hence in the emission/viewing geometry, and determine whether such changes are also related, e.g. to the pulsar age or other parameters.  As discussed in Mignani et al.\  (2015), very few $\gamma$-ray pulsars have accurately measured X-ray light curves  and even less (Crab, Vela, PSR\, B0540$-$69, PSR\, B0656+14, Geminga) have been also detected as optical pulsars  (see, e.g. Mignani 2010 for a summary). Expanding this sample is, then, crucial to build a general picture of the pulsar emission processes at different wavelengths.  Exploiting the larger number of known radio/$\gamma$-ray pulsars, a systematic comparison between the $\gamma$-ray and radio light curve profiles has been recently carried out by Pierbattista et al.\ (2016), showing the diagnostic power of this approach.

\subsection{The PSR\, J0633+0632 nebula}

\begin{figure}
\centering
{\includegraphics[height=8.5cm,angle=270,clip=]{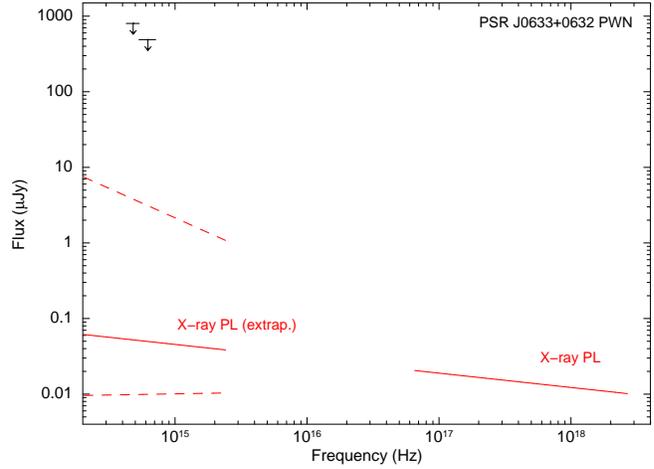}}
\caption{\label{pwn-spec} 
Extinction-corrected optical flux upper limits (black) for the PSR\, J0633+0632 PWN, compared with the extrapolation in the optical domain of its X-ray PL spectrum (in red). }
\end{figure}

Finally, we compared the extinction-corrected upper limit on the optical flux of the PSR\, J0633+0632 PWN  (Sectn.\ 2.1) with its unabsorbed 0.3--10 keV X-ray flux. This is  $F_{\rm X}^{pwn} = 2.92^{+0.79}_{-0.81} \times 10^{-13}$ erg cm$^{-2}$ s$^{-1}$ (Abdo et al.\ 2013), computed by fitting the PWN area with an ellipse of semimajor and semiminor axis  of 0\farcm58 and 0\farcm54, respectively,  oriented 130$^{\circ}$ due East (Marelli 2012).  We subtracted the flux contribution of the point-like X-ray source southwest of the pulsar position (Fig. \ref{pwn}), which is only spatially coincident with the PWN. The extinction-corrected  optical flux of the PWN in the $g'$ band is $F_{\rm opt}^{pwn} \la 9.8 \times 10^{-13}$ erg cm$^{-2}$ s$^{-1}$,  integrated over the same area as used to compute the PWN X-ray flux. As done for the pulsars, we assumed the most conservative value of the interstellar extinction.  This yields an optical--to--X-ray flux ratio of $ F_{\rm opt}/F_{\rm x} \la 4.6$.
PWNe have been detected both in the optical and X-rays around the Crab pulsar, PSR\, J0205+6449, PSR\, B0540$-$69, and  PSR\, J1124$-$5916. Our upper limit on the $ F_{\rm opt}/F_{\rm x}$ for the PSR\, J0633+0632 PWN is above the values obtained for the other PWNe, which are typically $\sim$0.02--0.04, apart from the Crab PWN which has an  $ F_{\rm opt}/F_{\rm x}\sim 2$  (Zharikov et al.\ 2008).
This means that, owing to the faintness of the PSR\, J0633+0632 PWN in the X rays, much deeper optical observations are needed to set similar constraints on its optical emission.
We also compared the extinction-corrected optical spectral flux upper limit  on the PWN in the $g'$ and $r'$-bands with the extrapolation of its X-ray spectrum in the optical domain. Like in Abdo et al.\  (2013), we used the best-fit spectral index of the PWN, $\Gamma_{\rm X}^{pwn} = 1.19^{+0.59}_{-0.22}$. 
The PWN SED is shown in Fig. \ref{pwn-spec}. As seen, we cannot rule the presence of a spectral break between the optical and the X-ray energy range.  A break in the optical/X-ray SED has been observed in other PWNe. For instance, the PWN around PSR\, B0540$-$69 features a clear break, with the optical fluxes being fainter than expected from the extrapolation of the X-ray PWN spectrum (Mignani et al.\ 2012). This is also the case for the PSR\, J1124$-$5916 PWN (Zharikov et al.\ 2008). A break in the opposite direction is observed in the SED of the PSR\, J1833$-$1034 PWN (Zajczyk et al.\ 2012), where the infrared fluxes (the PWN is not yet detected in the optical)  are about two orders of magnitude above the extrapolation of the PWN X-ray spectrum.  Only in the case of  the Crab and PSR\, J0205+6449 PWNe, the PWN spectrum is compatible with a single PL, extending from the X rays to the optical (Hester 2008; Shibanov et al.\ 2008).  Optical detections of more PWNe through dedicated observing campaigns can allow one to relate the differences in the SEDs to the characteristics of the PWN.

\section{Summary and conclusions}

Using data from the GTC, we carried out the deepest optical observations of the fields of the three moderately young  ($\sim 40$--60 kys old) $\gamma$-ray pulsars PSR\, J0248+6021, J0631+1036, and J0633+0632. The pulsars have not been detected down to $3 \sigma$ limits of $g'\sim27.3$,  $g'\sim27$,  and $g'\sim27.3$,  respectively. At the same time, we could not find evidence of extended optical emission associated with the faint X-ray PWN around PSR\, J0633+0632 (Ray et al.\ 2011).  Our limits on the $F_{\rm opt}/F_{\gamma}$ ratios are comparable with those of slightly younger, Vela-like pulsar, suggesting that pulsars in the age range 10--100 kyrs are quite similar in their optical and $\gamma$-ray emission and relatively less bright in the optical with respect to the $\gamma$-rays than the very young, Crab-like pulsar. In particular, our optical flux upper limits for PSR\, J0248+6021, J0631+1036, and J0633+0632 seem to support the idea that the fraction of the spin-down power converted into optical luminosity is much lower for pulsars  in the age range 10--100 kyrs than for Crab-like pulsars. Direct and precise distant measurements for the three pulsars discussed here will confirm this conclusion. Using archival \chan\ data we also searched for the unidentified X-ray counterpart to PSR\, J0248+6021 but we could not detect the pulsar down to a 0.3--10 keV flux limit $F_{\rm X} \sim 5 \times 10^{-14}$ erg cm$^{-2}$ s$^{-1}$, confirming the non-detection by Prinz \& Becker (2015), which improves the previous {\em Suzaku} limit (Abdo et al.\ 2013) by a factor of 20 and better constraints its $F_{\gamma}/F_{\rm X}$ ratio. With that computed for PSR\, J0631+1036, this limit indicates that these two radio-loud pulsars are more similar in their high-energy behaviour to the radio-quiet pulsars rather than to the bulk of the radio-loud ones. More X-ray observations of $\gamma$-ray pulsars are needed to keep progressing in the understanding of the similarities and differences between these two populations (Marelli et al.\  2011; 2015).
Although challenging, optical observations bring the tile required to complete the description of the multi-wavelength phenomenology of $\gamma$-ray pulsars. To this aim, X-ray observations are key to provide direct estimates (hopefully, as accurate as possible)  of the interstellar extinction, whereas radio parallax measurements are key to provide reliable distance estimates, at least for radio-loud pulsars.  Large uncertainties on both quantities dramatically impact on the optical follow-up  of objects as faint as neutron stars.

\section*{Acknowledgments}
We thank the anonymous referee for his/her careful revision of the manuscript and for the many useful suggestions. Based on observations made with the Gran Telescopio Canarias (GTC), installed in the Spanish Observatorio del Roque de los Muchachos of the Instituto de Astrof’sica de Canarias, in the island of La Palma. We thank Riccardo Scarpa (IAC) for useful advice in the data analysis. RPM acknowledges financial support from the project TECHE.it. CRA 1.05.06.04.01 cap 1.05.08 for the project "Studio multilunghezze dÕonda da stelle di neutroni con particolare riguardo alla emissione di altissima energia". 
NR acknowledges support via an NWO Vidi award. NR, DFT and EOW are supported by grants  SGR2014-1073 and AYA2015, and partially by the COST Action NewCOMPSTAR (MP1304). The work of MM was supported by the ASI-INAF contract I/037/12/0, art.22 L.240/2010 for the project $"$Calibrazione ed Analisi del satellite NuSTAR$"$.

\label{lastpage}

\end{document}